\documentclass[twocolumn,showpacs,superscriptaddress,aps,prc,10pt]{revtex4-1}
\usepackage[pdftex]{graphicx}
\usepackage[usenames,svgnames,table]{xcolor}
\usepackage[english]{babel}
\usepackage{ucs}
\usepackage[utf8x]{inputenc}
\usepackage{siunitx}
\usepackage[version=3]{mhchem}
\usepackage{multirow}
\newcommand{\garf}{\textsc{Garfield}}
\newcommand{\gemini}{\textsc{Gemini++}}
\newcommand\T{\rule{0pt}{2.6ex}}       
\newcommand\B{\rule[-1.2ex]{0pt}{0pt}} 
\begin{document}
\sisetup{exponent-product = \cdot,per-mode = symbol,range-phrase = --,list-units = single,range-units = single,separate-uncertainty = true,table-number-alignment = center}


\title{Charged particle decay of hot and rotating \ce{^{88}Mo} nuclei in fusion-evaporation reactions}

\author{S.~Valdré}
\email{valdre@fi.infn.it}
\affiliation{Dipartimento di Fisica, Università di Firenze, Italy}
\affiliation{INFN sezione di Firenze, I-50019 Sesto Fiorentino, Italy}

\author{S.~Piantelli}
\affiliation{INFN sezione di Firenze, I-50019 Sesto Fiorentino, Italy}

\author{G.~Casini}
\affiliation{INFN sezione di Firenze, I-50019 Sesto Fiorentino, Italy}

\author{S.~Barlini}
\affiliation{Dipartimento di Fisica, Università di Firenze, Italy}
\affiliation{INFN sezione di Firenze, I-50019 Sesto Fiorentino, Italy}

\author{S.~Carboni}
\affiliation{Dipartimento di Fisica, Università di Firenze, Italy}
\affiliation{INFN sezione di Firenze, I-50019 Sesto Fiorentino, Italy}

\author{M.~Ciema\l a}
\affiliation{Institute of Nuclear Physics Polish Academy of Sciences, 31-342 Krak\'ow, Poland}

\author{M.~Kmiecik}
\affiliation{Institute of Nuclear Physics Polish Academy of Sciences, 31-342 Krak\'ow, Poland}

\author{A.~Maj}
\affiliation{Institute of Nuclear Physics Polish Academy of Sciences, 31-342 Krak\'ow, Poland}

\author{K.~Mazurek}
\affiliation{Institute of Nuclear Physics Polish Academy of Sciences, 31-342 Krak\'ow, Poland}

\author{M.~Cinausero}
\affiliation{INFN, Laboratori Nazionali di Legnaro, I-35020 Legnaro, Italy}

\author{F.~Gramegna}
\affiliation{INFN, Laboratori Nazionali di Legnaro, I-35020 Legnaro, Italy}

\author{V.L.~Kravchuk}
\affiliation{National Research Centre ``Kurchatov Institute'', 123182 Moscow, Russia}

\author{L.~Morelli}
\affiliation{Dipartimento di Fisica e Astronomia, Università di Bologna and INFN sezione di Bologna, I-40127 Bologna, Italy}

\author{T.~Marchi}
\affiliation{INFN, Laboratori Nazionali di Legnaro, I-35020 Legnaro, Italy}

\author{G.~Baiocco}
\affiliation{Dipartimento di Fisica, Università di Pavia, Italy and INFN sezione di Pavia, I-27100 Pavia, Italy}

\author{L.~Bardelli}
\affiliation{Dipartimento di Fisica, Università di Firenze, Italy}
\affiliation{INFN sezione di Firenze, I-50019 Sesto Fiorentino, Italy}

\author{P.~Bednarczyk}
\affiliation{Institute of Nuclear Physics Polish Academy of Sciences, 31-342 Krak\'ow, Poland}

\author{G.~Benzoni}
\affiliation{INFN sezione di Milano, I-20133 Milano, Italy}

\author{M.~Bini}
\affiliation{Dipartimento di Fisica, Università di Firenze, Italy}
\affiliation{INFN sezione di Firenze, I-50019 Sesto Fiorentino, Italy}

\author{N.~Blasi}
\affiliation{INFN sezione di Milano, I-20133 Milano, Italy}

\author{A.~Bracco}
\affiliation{Dipartimento di Fisica, Università di Milano, I-20133 Milano, Italy}
\affiliation{INFN sezione di Milano, I-20133 Milano, Italy}

\author{S.~Brambilla}
\affiliation{INFN sezione di Milano, I-20133 Milano, Italy}

\author{M.~Bruno}
\affiliation{Dipartimento di Fisica e Astronomia, Università di Bologna and INFN sezione di Bologna, I-40127 Bologna, Italy}

\author{F.~Camera}
\affiliation{Dipartimento di Fisica, Università di Milano, I-20133 Milano, Italy}
\affiliation{INFN sezione di Milano, I-20133 Milano, Italy}

\author{A.~Chbihi}
\affiliation{Grand Accélérateur National d'Ions Lourds (GANIL), BP 55027, F-14076 Caen Cedex 5, France}

\author{A.~Corsi}
\affiliation{Dipartimento di Fisica, Università di Milano, I-20133 Milano, Italy}
\affiliation{INFN sezione di Milano, I-20133 Milano, Italy}

\author{F.C.L.~Crespi}
\affiliation{Dipartimento di Fisica, Università di Milano, I-20133 Milano, Italy}
\affiliation{INFN sezione di Milano, I-20133 Milano, Italy}

\author{M.~D'Agostino}
\affiliation{Dipartimento di Fisica e Astronomia, Università di Bologna and INFN sezione di Bologna, I-40127 Bologna, Italy}

\author{M.~Degerlier}
\affiliation{Nevsehir Haci Bektas Veli University, Science and Art Faculty, Physics Department, Turkey}

\author{D.~Fabris}
\affiliation{INFN, Sezione di Padova, Padova, Italy}

\author{B.~Fornal}
\affiliation{Institute of Nuclear Physics Polish Academy of Sciences, 31-342 Krak\'ow, Poland}

\author{A.~Giaz}
\affiliation{Dipartimento di Fisica, Università di Milano, I-20133 Milano, Italy}
\affiliation{INFN sezione di Milano, I-20133 Milano, Italy}

\author{M.~Krzysiek}
\affiliation{Institute of Nuclear Physics Polish Academy of Sciences, 31-342 Krak\'ow, Poland}

\author{S.~Leoni}
\affiliation{Dipartimento di Fisica, Università di Milano, I-20133 Milano, Italy}
\affiliation{INFN sezione di Milano, I-20133 Milano, Italy}

\author{M.~Matejska-Minda}
\affiliation{Institute of Nuclear Physics Polish Academy of Sciences, 31-342 Krak\'ow, Poland}
\affiliation{Heavy Ion Laboratory, University of Warsaw, 02-093 Warsaw, Poland}

\author{I.~Mazumdar}
\affiliation{Tata Institute of Fundamental Research, 400005 Mumbai, India}

\author{W.~M\c{e}czy\'nski}
\affiliation{Institute of Nuclear Physics Polish Academy of Sciences, 31-342 Krak\'ow, Poland}

\author{B.~Million}
\affiliation{INFN sezione di Milano, I-20133 Milano, Italy}

\author{D.~Montanari}
\affiliation{Dipartimento di Fisica, Università di Milano, I-20133 Milano, Italy}
\affiliation{INFN sezione di Milano, I-20133 Milano, Italy}

\author{S.~Myalski}
\affiliation{Institute of Nuclear Physics Polish Academy of Sciences, 31-342 Krak\'ow, Poland}

\author{R.~Nicolini}
\affiliation{Dipartimento di Fisica, Università di Milano, I-20133 Milano, Italy}
\affiliation{INFN sezione di Milano, I-20133 Milano, Italy}

\author{A.~Olmi}
\affiliation{INFN sezione di Firenze, I-50019 Sesto Fiorentino, Italy}

\author{G.~Pasquali}
\affiliation{Dipartimento di Fisica, Università di Firenze, Italy}
\affiliation{INFN sezione di Firenze, I-50019 Sesto Fiorentino, Italy}

\author{G.~Prete}
\affiliation{INFN, Laboratori Nazionali di Legnaro, I-35020 Legnaro, Italy}

\author{O.J.~Roberts}
\affiliation{University of York, Heslington, YO10 5DD York, UK}

\author{J.~Stycze\'n}
\affiliation{Institute of Nuclear Physics Polish Academy of Sciences, 31-342 Krak\'ow, Poland}

\author{B.~Szpak}
\affiliation{Institute of Nuclear Physics Polish Academy of Sciences, 31-342 Krak\'ow, Poland}

\author{B.~Wasilewska}
\affiliation{Institute of Nuclear Physics Polish Academy of Sciences, 31-342 Krak\'ow, Poland}

\author{O.~Wieland}
\affiliation{INFN sezione di Milano, I-20133 Milano, Italy}

\author{J.P.~Wieleczko}
\affiliation{Grand Accélérateur National d'Ions Lourds (GANIL), BP 55027, F-14076 Caen Cedex 5, France}

\author{M.~Zi\c{e}bli\'nski}
\affiliation{Institute of Nuclear Physics Polish Academy of Sciences, 31-342 Krak\'ow, Poland}

\date{\today}

\begin{abstract}
A study of fusion-evaporation and (partly) fusion-fission channels for the \ce{^{88}Mo} compound nucleus,
produced at different excitation energies in the reaction \ce{^{48}Ti + ^{40}Ca} at \SIlist{300;450;600}{MeV} beam energies,
is presented. Fusion-evaporation and fusion-fission cross sections have been extracted and compared with the existing systematics.
Experimental data concerning light charged particles have been compared with the prediction of the statistical model
in its implementation in the \gemini\ code, well suited even for high spin systems, in order to tune the main model
parameters in a mass region not abundantly covered by exclusive experimental data. Multiplicities for light charged
particles emitted in fusion evaporation events are also presented.
Some discrepancies with respect to the prediction of the statistical model have been found for forward emitted $\alpha$-particles;
they may be due both to pre-equilibrium emission and to reaction channels (such as Deep Inelastic Collisions, QuasiFission/QuasiFusion)
different from the compound nucleus formation.
\end{abstract}
\maketitle
\section{Introduction}
The study of the de-excitation of compound nuclei dates back to the beginning of the modern nuclear physics,
but it has found renewed interest in recent years
(for example, \cite{Roy12} for medium mass systems and
\cite{Morelli14-1,Morelli14-2,DiNitto12,Dey06} for very light systems)
thanks to the improvement of the experimental techniques, allowing for a cleaner selection of exclusive decay channels.
In this way it is possible to put stringent constraints on the ingredients of the statistical model,
commonly used to describe the decay of a compound nucleus at excitation energies below \SI{3}{MeV/nucleon}.
The hypothesis behind the use of this model is that compound nuclei are equilibrated systems,
whose decay is independent of their previous history (i.e. the way in which they were formed),
but depends on the statistical competition among all the open channels.
The decay width of each channel depends on the phase space available in the final state and on the amplitude of the transition matrix.
The tuning of the parameters of the statistical model on the basis of the experimental data obtained from compound nucleus reactions
is important not only for the topic of low energy fusion reactions, but also for other classes of nuclear processes.
In fact many kinds of reactions, from low (\SIrange{5}{10}{MeV/nucleon}) to high ($\gtrsim\SI{100}{MeV/nucleon}$) beam energies,
such as deep inelastic collisions, multifragmentation, spallation, etc.,
can be theoretically described by means of models of the interaction phase
(e.g. dynamical models such as BNV \cite{Baran05}, AMD\cite{Ono99}, etc.)
followed by a statistical afterburner for the hot products. Since in many cases the effects
related to the nuclear interaction phase under investigation are weak and blurred by the afterburner,
it is important to describe the statistical decay as well as possible in order to reduce the uncertainties on the dynamical stage.
A good control of the decay process is also mandatory in GDR (Giant Dipole Resonance) or DDR (Dynamical Dipole Resonance)
studies because the collective excitation features are extracted from  continuous spectra dominated by the statistical
gamma rays from the source, which must be therefore correctly described by models.

The present work deals with the investigation of fusion-evaporation and fusion-fission channels for the \ce{^{88}Mo}
compound nucleus at three different excitation energies, produced in the reaction \ce{^{48}Ti + ^{40}Ca}
at \SIlist{300;450;600}{MeV} beam energy. In particular, we compared the experimental energy spectra and multiplicities
of light charged particles with the prediction of \gemini\ code \cite{Charity10}, a widely used implementation of
the statistical model, in order to constrain the relevant parameters in a mass-energy-spin region in which not many
experimental data of exclusive type, with a clean selection of the fusion channel, are available.
Moreover, in this work we have also extracted the experimental cross sections for the fusion-evaporation and
the fusion-fission channels, obtaining values compatible with the existing systematics \cite{Eudes13}.

This experiment is part of a measurement campaign aimed at the investigation of
hot, rotating \ce{^{88}Mo} nucleus. The results concerning GDR gamma decay are presented in
\cite{Ciemala14,Ciemala15}. The \ce{^{88}Mo} nucleus, as other medium mass nuclei,
is a good candidate for such kind of study because it presents a significant fission barrier even at high rotational energy,
and it consequently has low fission probability also at high spin values \cite{Sanders91}.
The use of \gemini\ as statistical code to be compared with our experimental data is thus a good choice,
since this code is particularly well suited for the decay of high spin nuclei.

Among the data available in literature for the mass region $A<100$ we can cite, for example,
\cite{LaRana87}, where the inclusive decay of \ce{^{67}Ga} at \SI{91}{MeV} excitation energy has been investigated.
Proton and $\alpha$-particle spectra at some polar angles have been compared with the prediction of a statistical code;
a good agreement for $\alpha$-particle center of mass energy spectra is obtained if a deformation of the source is taken into account.
On the contrary, proton energy spectra are not well reproduced. Instead, the angular distributions fit
quite well with the prediction of the code for both protons and $\alpha$-particles, when a deformed source is used.

An inclusive study of two different reactions producing the same compound nucleus \ce{^{67}Ga} at
\SI{1.9}{MeV/nucleon} excitation energy is presented in \cite{Brown99}.
The authors found that it is not possible to describe both the energy spectra and the angular distributions
of $\alpha$-particles by means of a statistical code with a unique set of parameters;
moreover, proton energy spectra require reduced evaporation barriers to be well reproduced.

In \cite{Nebbia94} proton and $\alpha$ energy spectra in coincidence with evaporation residues have been
measured for the reaction \ce{^{32}S + ^{74}Ge} at different beam energies in the range \SIrange{5}{13}{MeV/nucleon}.
From the comparison with statistical model predictions, the reverse level density parameter $K=\frac{A}{a}$
is found to be substantially independent of the excitation energy of the system in the mass region around $A=100$.
The dependence of $K$ on the angular momentum has been investigated in \cite{Roy12} for
the compound nuclei \ce{^{97}Tc} and \ce{^{62}Zn} at \SI{36}{MeV} excitation energy, comparing the energy spectra of neutrons,
protons and $\alpha$-particles with the prediction of the \textsc{Cacarizo} code (Monte Carlo version
of the \textsc{Cascade} statistical model code \cite{Puhlhofer77});
the angular momentum selection was performed by means of the $\gamma$ multiplicity.
The authors found that the deformation of the source (taken into account by means of the deformability parameters included in the code)
has no effect on neutrons and protons energy spectra, while it significantly modifies those of
$\alpha$-particles for the lighter compound nucleus. The $K$ values extracted from proton,
neutron and $\alpha$-particle spectra are different, but they all show a decrease when the
angular momentum of the source increases (in the range \SIrange{13}{22}{\planckbar}).

In \cite{Bhattacharya01} $\alpha$ energy spectra detected in coincidence with Evaporation Residues (ER)
from \ce{^{56}Ni} CN produced in the symmetric reaction \ce{^{28}Si + ^{28}Si} at \SI{112}{MeV}
beam energy are compared with the prediction of \textsc{Cacarizo}.
The experimental energy spectra have been well reproduced by introducing a high degree of deformation
(corresponding to a quadrupole deformation parameter equal to 0.5) which modifies the
effective moment of inertia of nuclei, thus lowering the yrast line. In this way the decay chain becomes longer
and $\alpha$-particles are emitted later in the cascade, with smaller average kinetic energy.
The authors claimed also that deformation at high spin values is particularly favoured if the entrance channel is mass symmetric.
Note that the spin range explored by this system is lower than in the present \ce{^{88}Mo} case;
in fact the critical angular momentum for fusion was \SI{34}{\planckbar} in \cite{Bhattacharya01},
while it is \SI{64}{\planckbar} in our case.

In \cite{Kildir92} a detailed study of the transmission coefficients of the statistical model,
based on the energy spectra of hydrogen isotopes and $\alpha$-particles evaporated by compound nuclei of \ce{^{96}Ru}
at \SI{1.2}{MeV/nucleon} excitation energy, is presented. $\alpha$-particle spectra are found to be in good agreement
with the prediction of the statistical model (\textsc{Cascade} code) if reduced barriers with respect
to the prescription of the optical model are used. Moreover, the level density has to be enhanced,
indicating the onset of deformations at high spins (kept into account by lowering the yrast line as done in \cite{Bhattacharya01}).
For protons, on the contrary, the authors explain the unsatisfactory agreement between experimental and simulated energy
spectra with the need of taking into account dynamical effects.

In \cite{Brekiesz07} and \cite{Kmiecik07} spectra of $\alpha$-particles emitted from \ce{^{46}Ti} CN with \SI{85}{MeV}
excitation energy and critical angular momentum of \SI{35}{\planckbar} are discussed and compared with the prediction
of the already cited \textsc{Cacarizo} code. Various parametrizations of the yrast line implying different degrees
of deformation of the CN have been tested, obtaining reasonable agreement with the experimental data.
In particular, energy spectra of $\alpha$-particles detected in coincidence with a residue of $Z=20$
(corresponding to chains in which only one $\alpha$-particle is emitted) can be reproduced only if an extremely
high value of the deformation parameter is used. The authors interpret this fact as due either to the presence
of a dynamical hyperdeformed state or to the pre-equilibrium emission from a dinuclear system not completely fused.
This study is coupled to the investigation of the GDR, confirming the presence of a strongly deformed structure.

In summary, all these  studies demonstrate that the decay features of light systems with high spin (up to $\sim\SI{60}{\planckbar}$)
are a difficult task for available statistical codes.
In fact, a single parametrization is often not able to reproduce both proton and $\alpha$-particle spectra.
Moreover, in many cases a high degree of deformation is necessary to obtain reasonable results.
We anticipate that also our results show that the statistical model with different assumptions of
the source parameters is not able to reproduce all the characteristics of the emitted particles,
in particular of the $\alpha$ particles, with increasing bombarding energy.

\section{Experimental setup and ion identification}
\label{sec:experiment}
\ce{^{88}Mo} compound nuclei have been produced at three
different excitation energies (\SIlist{1.4;2.2;3.0}{MeV/nucleon}) by means of fusion reactions performed
at the INFN \textit{Laboratori Nazionali di Legnaro} (LNL, Italy).
Pulsed beams of \ce{^{48}Ti} at the three bombarding energies (\SIlist{300;450;600}{MeV})
impinged on a metallic \ce{^{40}Ca} target (\SI{500}{\micro\gram\per\square\centi\metre} thickness)
sandwiched between two very thin C foils (\SI{15}{\micro\gram\per\square\centi\metre}) to prevent prompt oxidation.
Typical beam currents of \SIrange{0.5}{1}{pnA} were used.
In order to study the GDR evolution with excitation energy and spin in
molybdenum nuclei, a composite apparatus was used.
A group of 8 \ce{BaF2} scintillators (\textsc{Hector} setup \cite{Giaz14,Maj94,Ciemala14}) for gamma rays, covering
backward  laboratory angles, was coupled with a large acceptance
detector for charged reaction products. Here we describe only this
latter part of the setup, while details on the gamma array can be found in \cite{Ciemala14,Ciemala15}.

The heavy products, in particular evaporation residues from the fusion
reactions, were detected by an array of 48 triple-phoswiches of the
\textsc{Fiasco} setup \cite{Bini03} mounted in six matrices, each with eight
detectors, in an axially symmetric configuration around the beam
direction. The phoswiches featured two plastic layers
(made of a fast \SI{180}{\micro\metre}, and a slower \SI{5}{mm} scintillators)
followed by a \ce{CsI(Tl)} crystal \SI{4}{cm} thick, thus presenting a wide dynamic range for pulse shape analysis.
With respect to their original use, these detectors
were now equipped with digital electronics \cite{Pasquali07}, purposely developed
by the collaboration. The anode current pulses from the photomultipliers were digitized and the
relevant  information (time mark, pulse shape and energy-related variables) was extracted.
In practice, for each phoswich one obtains, besides a time mark for the time-of-flight (ToF),
three energy-related variables (henceforth gateA, gateB and gateC);
gateC corresponds to the light emitted by the \ce{CsI(Tl)} with the longest time constant;
gateB includes the light of the second plastic layer and part of the light emitted by the \ce{CsI(Tl)};
gateA includes the fastest light components from all the three scintillator layers.

The phoswiches covered the polar region from around \ang{5} to \ang{25};
32 of them were located in a wall configuration from \ang{5} to \ang{13} with a significant efficiency for ER
detection while the remaining 16 scintillators were placed in two side arms, mainly to detect fission fragments.
Thanks to the first fast plastic layer and to the large distance from the target (\SI{1.6}{m}), the
phoswiches permitted velocity measurement of the ejectiles from $Z=1$ up to the ER with good time resolution (of the order of ns). The charge identification, only possible
for the products punching-through the first thin layer,
has been obtained via digital pulse shape methods from protons to $Z$ close to the beam atomic number.
The upper limit in charge depends on the particular telescope and on the beam energy, and it never exceeds $Z = 18$.
Hydrogen isotopes have been identified also by means of the fast-slow technique in \ce{CsI(Tl)}.
Heavy fusion residues and heavy fission fragments, stopped in the first plastic layer, were detected without charge
identification. It has to be noted that the time mark given by the phoswich telescopes by means of the digital technique
(based on the numerical implementation of the Constant Fraction Discrimination method, used for the first time during this experiment)
was progressively delayed when the kinetic energy of the particles increases and thus the energy deposited
in the slow plastic layer and in \ce{CsI(Tl)} grows. This systematic distortion (walk in the time mark detection)
was associated with a too small amount of digitized samples which were recorded for each phoswich signal and could
not be corrected in the off line analysis. As a consequence, in the following, we discuss velocity spectra for
heavy fragments which are stopped in the first layer, while for light products, entering also in the successive
layers of the telescopes, we are able to rely only on the particle multiplicities; in fact we judge the velocity
(and thus the energy) not trustworthy.

The forward chamber of \garf\ \cite{Bruno13}, covering polar angles between
\ang{29.5} and \ang{83} is able to detect light charged particles (LCPs) and intermediate mass fragments (IMFs).
\garf\ is based on $\varDelta E\mbox{(gas)}-E\mbox{(\ce{CsI(Tl)})}$
modules, equipped with digital electronics and its features and
performances are extensively described elsewhere \cite{Bruno13}.
Very briefly, \garf\ consists of 192 $\varDelta E\mbox{(gas)}-E\mbox{(\ce{CsI(Tl)})}$
telescopes. The gas volume is unique for all the modules and it is
filled with flowing \ce{CF4} at a pressure around \SI{50}{mbar}. The collecting
anodes of the $\varDelta E$ stage, based on metal/glass microstrip
technology, provide a moderate gas multiplication. While preserving the
linear response with deposited energy, the internal gain allows for
detection and identification from light charged particles to intermediate mass
fragments (typically $Z\approx 10-12$). The 192 $\varDelta E\mbox{(gas)}-E\mbox{(\ce{CsI(Tl)})}$
telescopes are organized in four polar rings corresponding to four \ce{CsI(Tl)} shapes;
each ring consists of 48 effective sectors, with an azimuthal granularity of \ang{7.5}.
In this experiment the mechanical constraints, related to the coupling
of the \ce{BaF2} with \garf , imposed a change in the target holder. As a
result the backward ring of \garf\ (from \ang{67} to \ang{83}) was
affected by the target shadowing and has been discarded in the present analysis.

As already said, for each telescope the $\varDelta E-E$ method
allows to identify charged particles from $Z=2$ to $Z\approx 10-12$,
measuring also their energy. Moreover, light charged particles ($Z=1,2$) were
isotopically resolved via pulse shape analysis in the
\ce{CsI(Tl)} crystals \cite{Morelli10}. The energy calibration of \ce{CsI(Tl)} has been
carried out by exploiting the many reference points
collected in various measurements of elastic diffusion with low energy beams. The overall
uncertainty on the energy measured by the \ce{CsI(Tl)} is $\approx\;$\SIrange{3}{4}{\%}.
For kinematic reasons, heavy residues cannot reach the angles of the
\garf\ setup; as a consequence, \garf\ was able to identify (and
measure the energy of) all the charged products 
hitting its active surface.

Finally, a small plastic scintillator was located at $\vartheta\approx\ang{2}$, well below the laboratory grazing
angle for all the investigated reactions (the most forward one is
$\vartheta_{\mathrm{graz}}=\ang{6.9}$ for the \SI{600}{MeV} reaction).
It covered a small solid angle (\SI{3.6e-5}{sr}) and has been used to count
elastically scattered ions for absolute cross section normalization.

\section{Experimental results}
\label{sec:results}
As stated in the introduction, the topic of this work consists in the investigation of the decay
of the excited compound nucleus formed in the \ce{^{48}Ti + ^{40}Ca} reaction at three beam energies
and the comparison with the prediction of the statistical code.
The main reaction parameters for the investigated systems are reported in
Table~\ref{tab:param}. In the columns labelled $l^\mathrm{B_f=0}_\mathrm{RLDM}$ and $l^\mathrm{B_f=0}_\mathrm{Sierk}$ two estimations of the maximum angular momentum for the existence of the compound nucleus are reported. The former is calculated  in the framework of the rotating liquid drop model (RLDM) of \cite{Cohen74}, while the latter  is obtained according to the modified rotating liquid drop model of Sierk \cite{Sierk86}. At \SI{300}{MeV}, where the grazing angular momentum $l_\mathrm{graz}$ (obtained from \cite{Gupta84}) does not exceed by a large amount $l^\mathrm{Bf=0}_\mathrm{Sierk/RLDM}$, the complete fusion represents the dominant reaction channel. The formed compound nucleus has a broad spin distribution \cite{Ciemala15} and it decays either by evaporation (fusion-evaporation (FE) process) or by fission
(fusion-fission (FF) process, possibly followed by the evaporative decay from the fission fragments).
The two fission products can be of similar mass (symmetric fission) or may be formed by a lighter and a heavier fragment (asymmetric fission). The FF channel becomes more and more significant with respect to the
FE process when the angular momentum of the compound nucleus increases, up to the $l^\mathrm{Bf=0}_\mathrm{Sierk/RLDM}$ value,
beyond which the system becomes unstable against fission, because the fission barrier vanishes.
Beyond $l^\mathrm{Bf=0}_\mathrm{Sierk/RLDM}$ fusion is prevented and other reaction mechanisms take place;
in particular the deep inelastic process is the most important, but also the quasi fusion / quasi fission mode
(in which the system does not go through a completely equilibrated fused phase) can be present.
Those mechanisms, which are supposed to increase with beam energy, produce two heavy residues in the exit channel
(for the deep inelastic process, the quasi projectile QP and the quasi target QT), similar to symmetric fission.
The contribution of such binary channels represents a background for the true FF events which can be barely disentangled,
because the populated phase space regions are partly overlapping. These mechanisms affect mainly the symmetric fission
case because the two reaction partners are quite similar in terms of mass.

\begin{table*}[htbp]
\caption{Main reaction parameters.
	$E_\mathrm{b}$ is the beam energy; $E_\mathrm{CM}$ is the available energy in the center of mass,
	after removing the energy lost in half target; $\vartheta_\mathrm{graz}$ and $l_\mathrm{graz}$ are,
	respectively, the grazing angle and the grazing angular momentum, calculated according to \cite{Gupta84}; $l^\mathrm{B_f=0}_\mathrm{RLDM}$ is the critical angular momentum beyond which the fission barrier vanishes, calculated according to \cite{Cohen74};
	$l^\mathrm{B_f=0}_\mathrm{Sierk}$ is the critical angular momentum beyond which the fission barrier vanishes, calculated according to  \cite{Sierk86};  $E^{*}_\mathrm{CN}$ is the excitation energy
	of the compound nucleus, calculated according to the formula $E^*_\mathrm{CN}=E_\mathrm{CM}+Q$,
	where $Q$ is the Q-value for fusion; $E^{*}_{CN}/A$ is the excitation energy per nucleon of the compound nucleus; $\sigma_\mathrm{R}$
	is the reaction cross section calculated according to \cite{Gupta84}.
}
\centering
\begin{tabular}{ccccccccc}
\toprule
\T {~$\boldsymbol{E}_\mathbf{b}$~} & {~$\boldsymbol{E}_\mathbf{CM}$~} & {~$\boldsymbol{\vartheta}_\mathbf{graz}$~} &%
{~$\boldsymbol{l}_\mathbf{graz}$~} & {~$\boldsymbol{l}^\mathbf{B_f=0}_\mathbf{RLDM}$~} & {~$\boldsymbol{l}^\mathbf{B_f=0}_\mathbf{Sierk}$~}&%
{~$\boldsymbol{E}^*_\mathbf{CN}$~} & {~$\boldsymbol{E}^*_\mathbf{CN}/\boldsymbol{A}$~} & {~$\boldsymbol{\sigma}_\mathbf{R}$~}\\
\T\B {~[\si{MeV}]~} & {~[\si{MeV}]~} & {~[\textdegree]~} & {~[\si{\planckbar}]~} & {~[\si{\planckbar}]~} & {~[\si{\planckbar}]~} &%
{~[\si{MeV}]~} & {~$\left[{\frac{\si{MeV}}{\si{nucleon}}}\right]$~} & {~[mbarn]~}\\
\colrule
\T 300 & 134.7 & 15.8 &  91 & 79 & 64 & 123.8 & 1.4 & 1863\\
   450 & 203.0 &  9.5 & 124 & 79 & 64 & 192.4 & 2.2 & 2268 \\
\B 600 & 271.0 &  6.9 & 149 & 79 & 64 & 260.7 & 3.0 & 2446\\
\botrule
\end{tabular}
\label{tab:param}
\end{table*}

\subsection{Selection of fusion events}
\begin{figure*}[htbp]
\centering
\includegraphics[width=\textwidth] {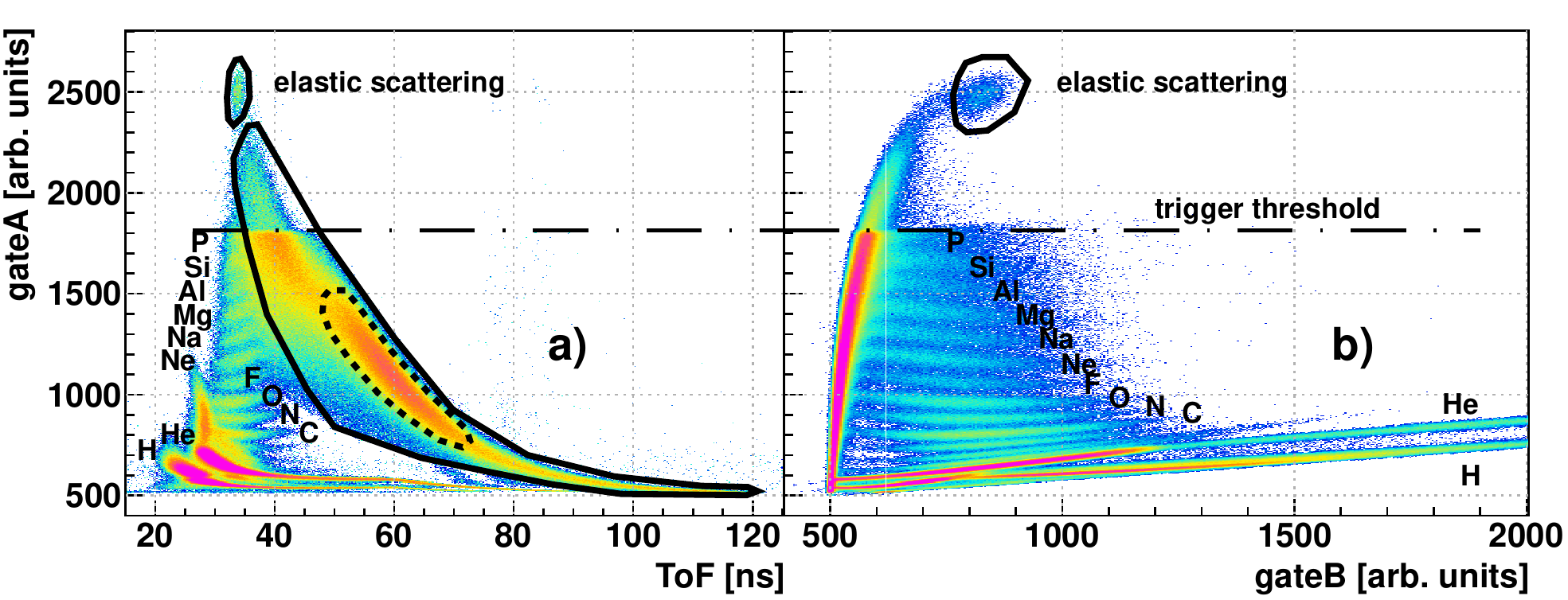}
\caption{(Color online) (a): correlation between gateA (see text) and the time of flight for a phoswich telescope
	at $\vartheta\simeq\ang{8.3}$; (b): correlation between gateA and gateB for the same phoswich telescope.
	Areal gates for particle identification are shown in the left panel and are explained in the text.
	Data refer to the reaction at \SI{600}{MeV}.
}
\label{fig:gates}
\end{figure*}

In order to properly select FE and FF events, a first pre-sorting was done at the trigger level.
Indeed, the main trigger was the coincidence of a hit in a phoswich and a hit in \garf .
Considering the phoswich and \garf\ acceptances, this condition strongly selects FE events. In order to
better suppress undesired events (i.e. elastic scattering and
binary peripheral reactions) only detector signals above the noise level but below a
given threshold (corresponding to gateA values around \SI{2000}{ch}, thick horizontal line in Fig.~\ref{fig:gates}) were allowed to
start the data acquisition. In Fig.~\ref{fig:gates}(a) we show a typical gateA vs. ToF correlation,
which is the main criterion for the off-line event selection, while in part (b) of the picture the gateA vs. gateB correlation is presented.
The plots refer to the reaction at \SI{600}{MeV}.
In the gateA vs. ToF correlation three different regions can be identified: ions stopped in the first
thin plastic scintillator layer (hyperbolic branch at $\mathrm{ToF}>\SI{40}{ns}$), for
which the charge identification is not possible; IMFs and LCPs stopped in the second scintillator layer
(spots at different gateA values), for which the charge identification is possible and finally LCPs
(only protons for the reactions at \SI{300}{MeV}) entering the \ce{CsI(Tl)} layer (sudden increase of the gateA value).
Ions stopped in the first scintillator layer produce the almost vertical and unresolved branch in the gateA vs. gateB
correlation, while those punching through the fast plastic collapse on the almost horizontal lines, corresponding to the various $Z$,
visible in the picture. LPCs entering the \ce{CsI(Tl)} give rise to the discontinuity in the lower part of the gateA vs. gateB correlation.

FE events are selected by requiring that only one
fragment in the phoswich wall be inside the inner dashed-line areal gate
drawn on the gateA vs. ToF correlation in the region of unresolved heavy products.
In the same figure one can clearly see the yield drop above the
trigger upper level. The island of events at the highest gateA values
corresponds to residual elastic scattered ions and has been used for ToF calibration.

The fission of nuclei with $A\approx 100$, below the Businaro-Gallone point,  mainly produces two
fragments with different sizes. Therefore we expect that in our
system the asymmetric splits prevail, as reported in \cite{Jing99} for molybdenum isotopes.

Candidates for symmetric FF events have been selected requiring the coincidence of two heavy products
in the phoswich wall falling inside the outer full-line areal gate drawn on the left side of Fig.~\ref{fig:gates};
instead candidates for asymmetric FF events have been selected by the coincidence of a heavy product in the phoswich wall
(inside the same full-line areal gate) and an IMF in whatever detector section.
Moreover, candidate fission fragments must have their center of mass velocity collinear and their relative velocity
compatible with the fission systematics \cite{Viola65,Viola85}. For kinematical reasons,
two heavy fragments (symmetric fission) cannot reach \garf , while only the lighter fragment
from asymmetric splits can be detected in \garf . Whatever the emission angle, the IMF is the only partner of the fission
process which can be identified in charge. Of course, asymmetric fission events in which the light partner is missed produce
a background of incomplete events for the selected FE event set. This background has been estimated, as described later.

\begin{figure*}[htbp]
\centering
\begin{tabular}{cc}
\includegraphics[width=\columnwidth] {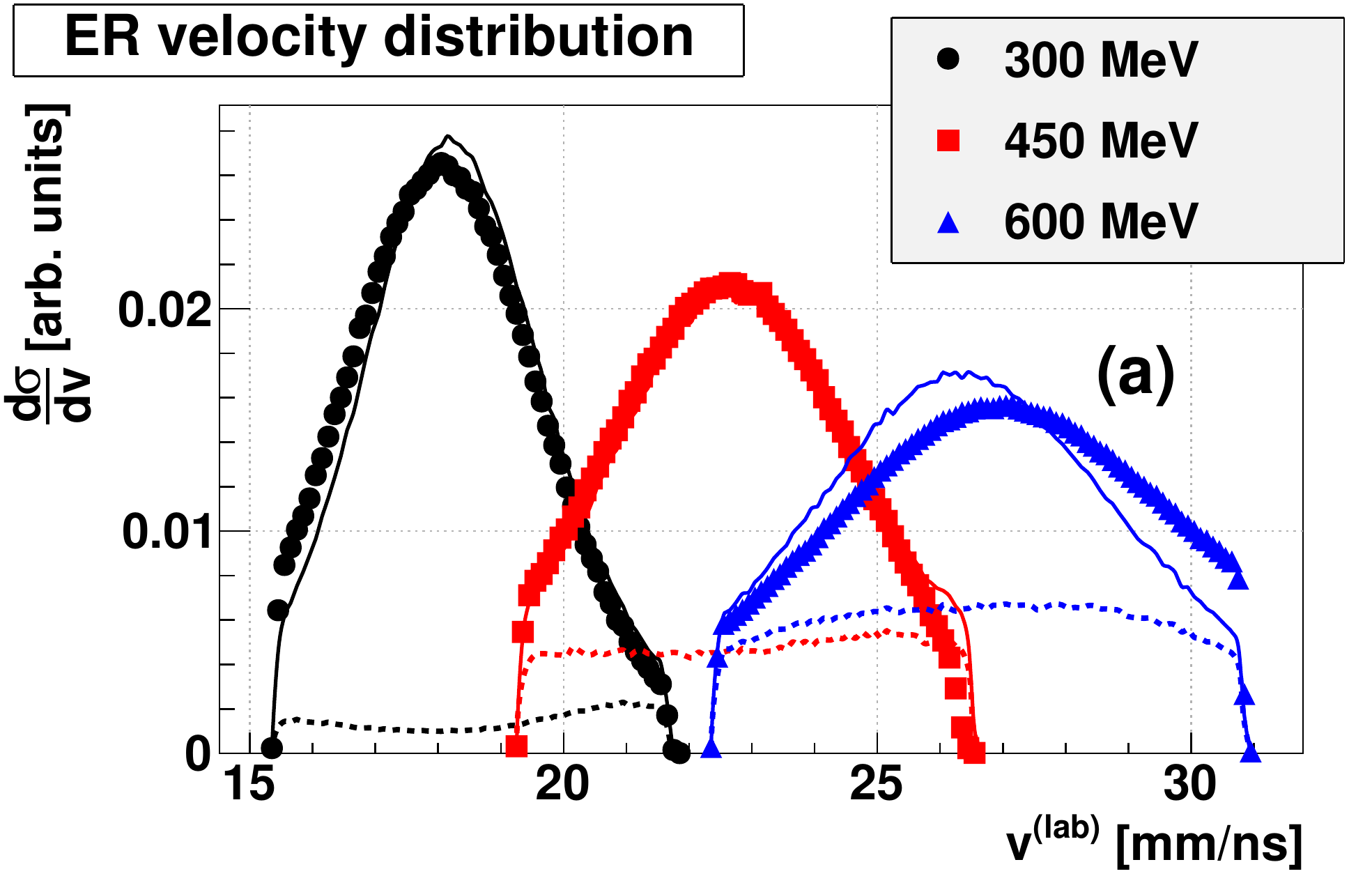} & \includegraphics[width=\columnwidth] {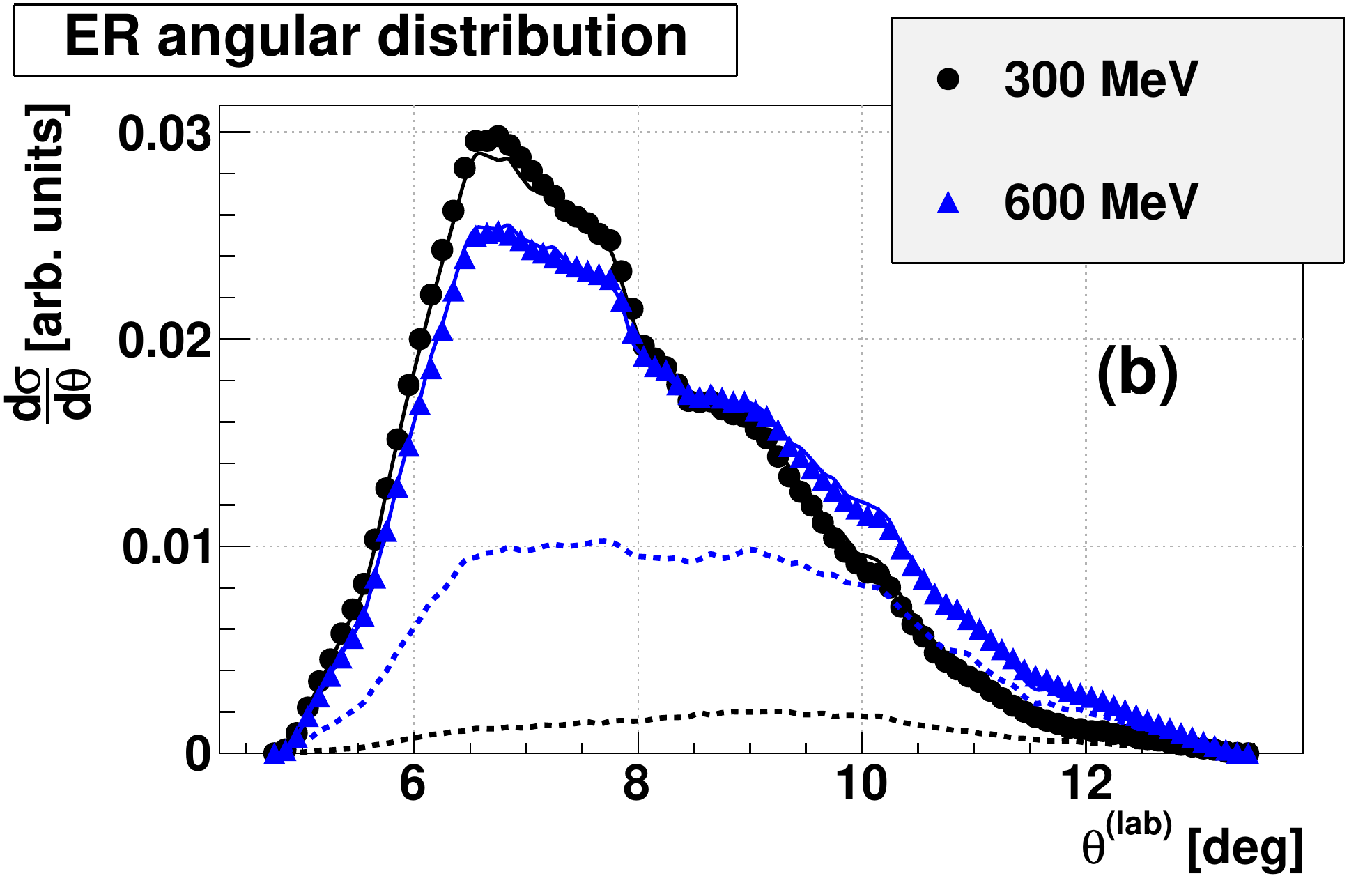}
\end{tabular}
\caption{
	(Color online) (a): ER lab velocity distribution at the three bombarding energies
	(black: \SI{300}{MeV}, red: \SI{450}{MeV}, blue: \SI{600}{MeV}); symbols are the experimental data, while
	continuous lines correspond to \gemini\ simulation
	with RLDM yrast, RLDM fission barrier, $w=\SI{1.0}{fm}$, $\tau_\mathrm{d}=\SI{0}{zs}$ (see text).
	Part (b): ER lab angular distribution at \SI{300}{MeV} (black) and \SI{600}{MeV} (blue);
	symbols are experimental data, while continuous lines correspond to \gemini\ simulation.
	In both pictures dotted lines correspond to simulated fission events in which the light partner
	was lost and the heavy one was erroneously identified as FE. All spectra are normalized to their integral,
	except for dotted lines (for which the same scaling factor of continuous curves has been used).
}
\label{fig:res_all}
\end{figure*}

\subsection{Use of the simulation}
We first consider the comparison of the experimental observables with \gemini\ in the reaction at \SI{300}{MeV},
where the assumption of the formation of the CN following complete fusion is most reliable; in particular,
we compared experimental energy spectra and multiplicities of protons and $\alpha$-particles
with the prediction of the code for the decay chain of a \ce{^{88}Mo} source with the proper excitation energy
(\SI{1.4}{MeV/nucleon}) and a triangular spin distribution up to the $l_\mathrm{crit}$ value.
In order to perform a correct comparison, the simulated events were filtered with a software replica of our setup.
The main code parameters were tuned in order to obtain the best overall agreement between experimental results and simulated data.
In particular, referring to the parameters introduced in \cite{Charity10}, we tried to tune the level density parameter
$\tilde{a}_\mathrm{eff}$ (acting, in particular, on the parameter $k_0$ of equation (15) of reference \cite{Charity10}),
the spread of the LCP Coulomb barrier taken into account by a parameter $w$ in the transmission coefficients,
the yrast parametrization $E_\mathrm{Y}$, the time delay for fission $\tau_\mathrm{d}$ and the parametrization
of the fission barrier $B_\mathrm{f}$ as a function of the total angular momentum $J$. According to \cite{Charity10},
the level density parameter rules the slope of the exponential tail of the energy spectra while $w$
controls their shape close to the Coulomb barrier. $E_\mathrm{Y}$ is particularly significant for light nuclei
with small moment of inertia; it affects the emission of $\alpha$ particles which remove large amount of spin and has slight influence on protons.
The $\tau_\mathrm{d}$ parameter and the shape of $B_\mathrm{f}$ can influence the occurrence
of the fission process and its competition with particle decay.
The model parameters have been tuned mainly looking at LCP spectra for FE events.

Once a reasonable parameter set has been fixed, we can use simulated events to reliably evaluate the
detector efficiency and thus estimate the $4\pi$-corrected particle multiplicities and the absolute
cross sections at all the measured beam energies.

\subsection{Fusion-Evaporation and Fusion-Fission}
According to the \gemini\ simulation, the detection efficiency for the evaporation residues (ER)
in FE events has been evaluated to be within \SI{10}{\%} and \SI{13}{\%} (increasing with the beam energy)
with the parametrization giving the best agreement between experimental and simulated LCP spectra (see in the following).
In Fig.~\ref{fig:res_all}(a) the experimental lab velocity distribution for the ER (full symbols)
is compared with the simulated one (filtered with a software replica of our setup, continuous lines) for
the three different beam energies; spectra are normalized to their integral.
Simulated and experimental data are generally well matched, with the partial exception of the \SI{600}{MeV} case;
as a consequence, we can argue that a residual pollution of other reaction mechanisms (not included in the simulation)
is present for the highest beam energy. On part (b) of the same figure the experimental lab angular distribution of the
ER (full symbols) for the reactions at \SI{300}{MeV} and \SI{600}{MeV} compared with the \gemini\ simulation (continuous lines)
is plotted. A very good agreement is obtained for this observable.
In both sides of Fig.~\ref{fig:res_all} dotted lines correspond to a fission background of ER events where
one of the fission products (the lighter one) was lost and the other one fell inside the ER detection gate
(dashed contour in Fig.~\ref{fig:gates}). In this case 
the scaling factor is the same used for the continuous curves. These events show an almost flat angular distribution and
they constitute a background of incompletely detected events, present both in simulated and experimental data.
According to the \gemini\ calculation, their amount is about \SI{9}{\%} of ER events at \SI{300}{MeV}, it rises to about \SI{34}{\%}
of ER at \SI{450}{MeV} and it is about \SI{50}{\%} of the total ER at the highest beam energy.

\begin{figure*}[htpb]
\centering
\begin{tabular}{cc}
\includegraphics[width=\columnwidth] {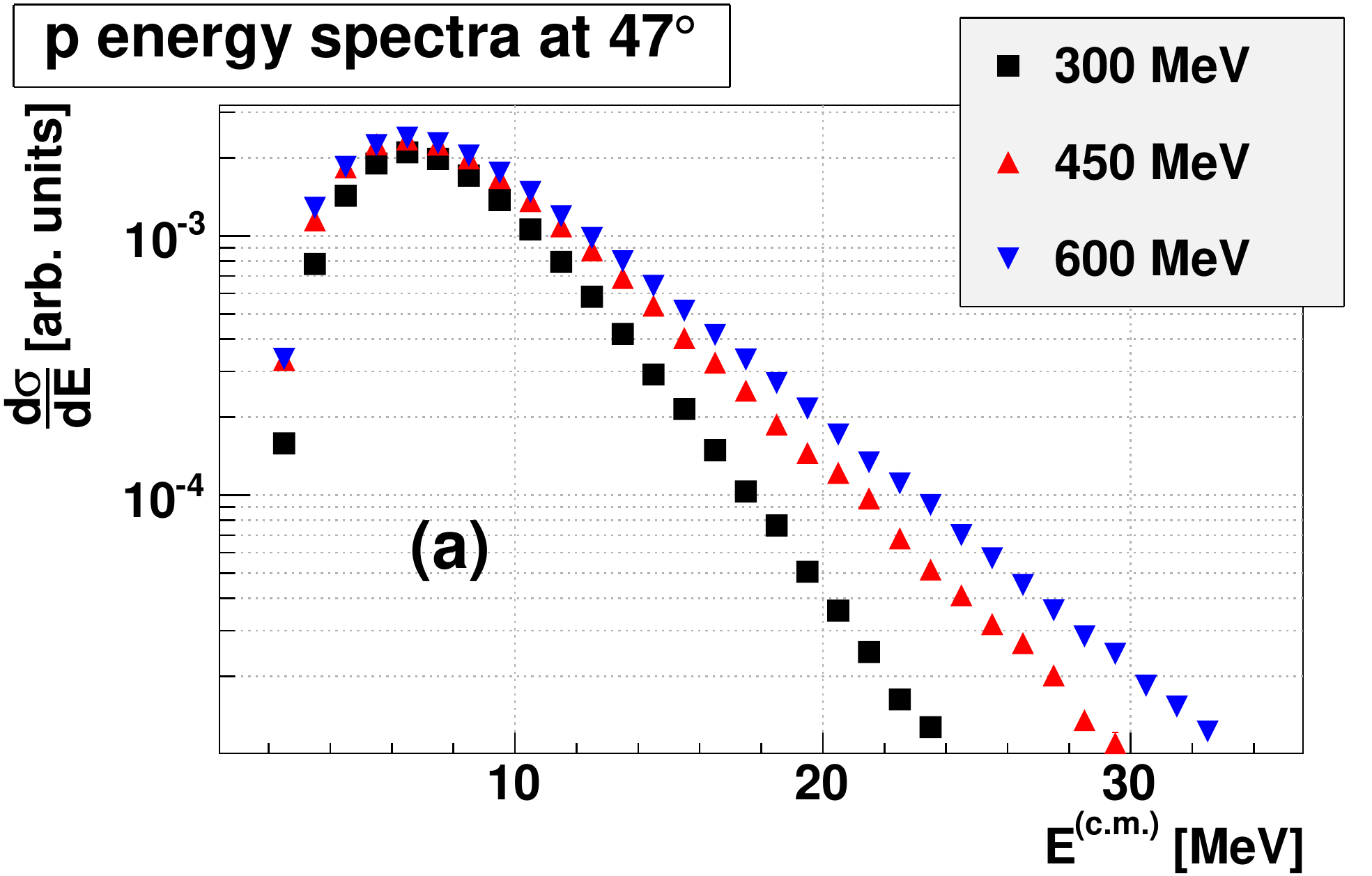} & \includegraphics[width=\columnwidth] {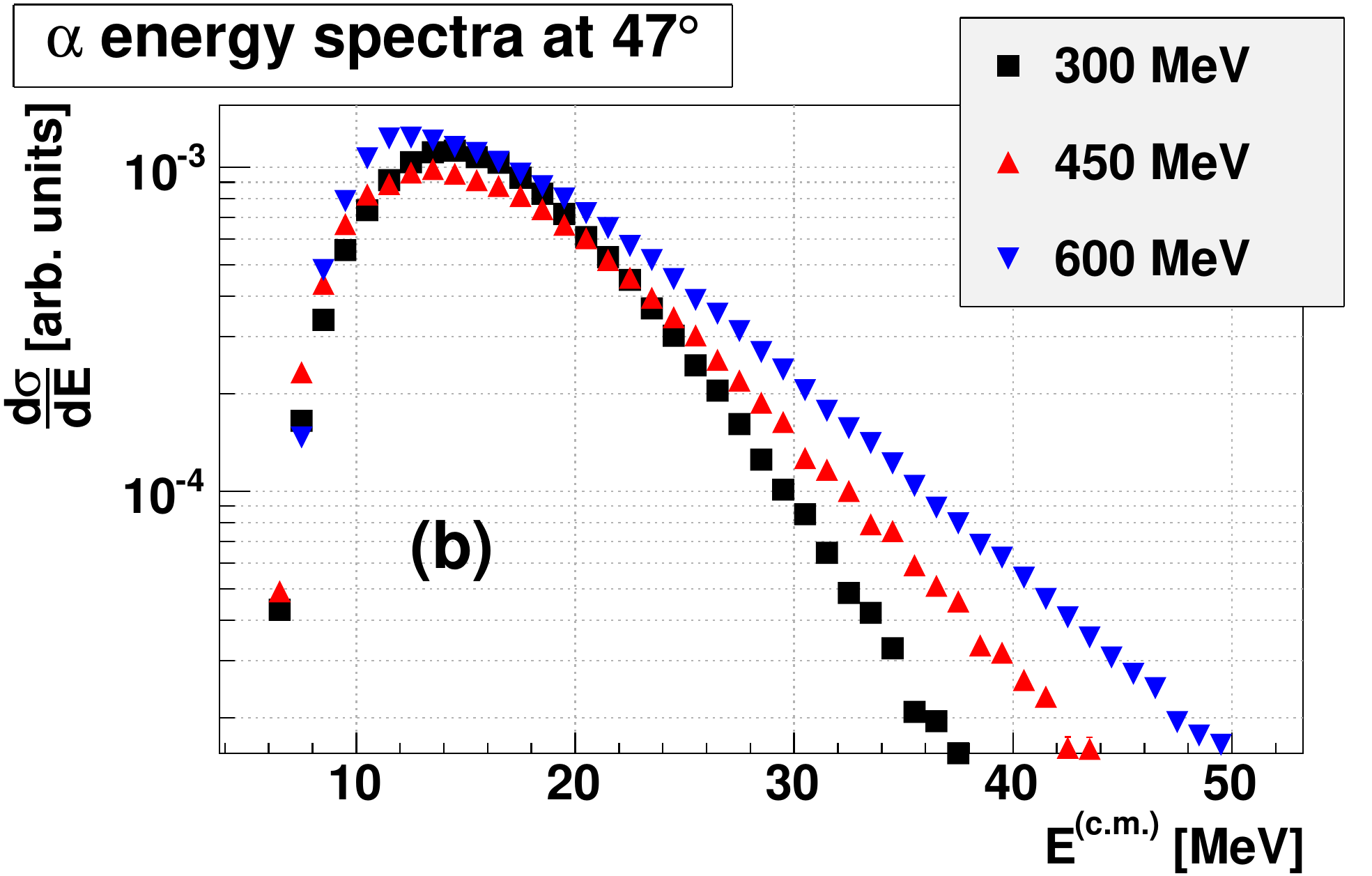}
\end{tabular}
\caption{
	(Color online) (a): experimental center of mass energy spectra for protons in the reaction at \SI{300}{MeV} (black squares),
	at \SI{450}{MeV} (red up triangles) and at \SI{600}{MeV} (blue down triangles),
	detected in the \garf\ ring with polar angles ranging from \ang{41} to \ang{52}.
	The spectra are normalized to the number of evaporation residues. (b): same for $\alpha$-particles.
}
\label{fig:exp}
\end{figure*}

Protons and $\alpha$-particles represent the dominant contribution in the decay chain of ER in FE events.
The experimental center of mass energy spectra of protons and $\alpha$-particles at all the investigated beam energies
are presented in part (a) and (b) of Fig.~\ref{fig:exp}, respectively; the spectra are normalized to the number of ER and refer
to particles detected in the \garf\ ring covering the polar angle between \ang{41} and \ang{52}.

As expected for the decay of a hot nuclear source, the spectra have Maxwellian shape,
with the apparent temperature (given by the inverse slope of the high energy tail) increasing with the beam energy,
as the excitation energy of the CN increases.

\begin{figure*}[htpb]
\centering
\begin{tabular}{cc}
\includegraphics[width=\columnwidth] {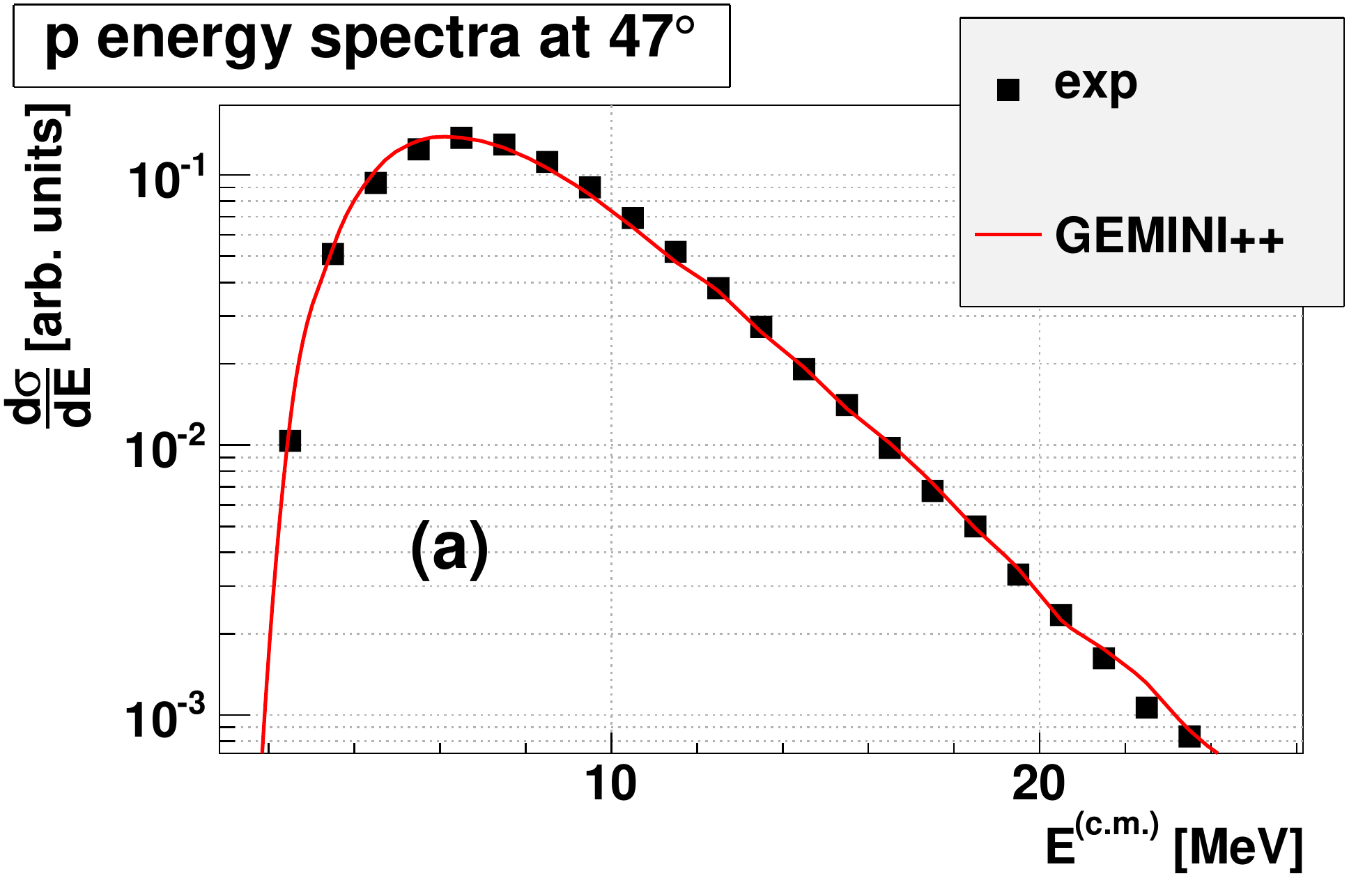} & \includegraphics[width=\columnwidth] {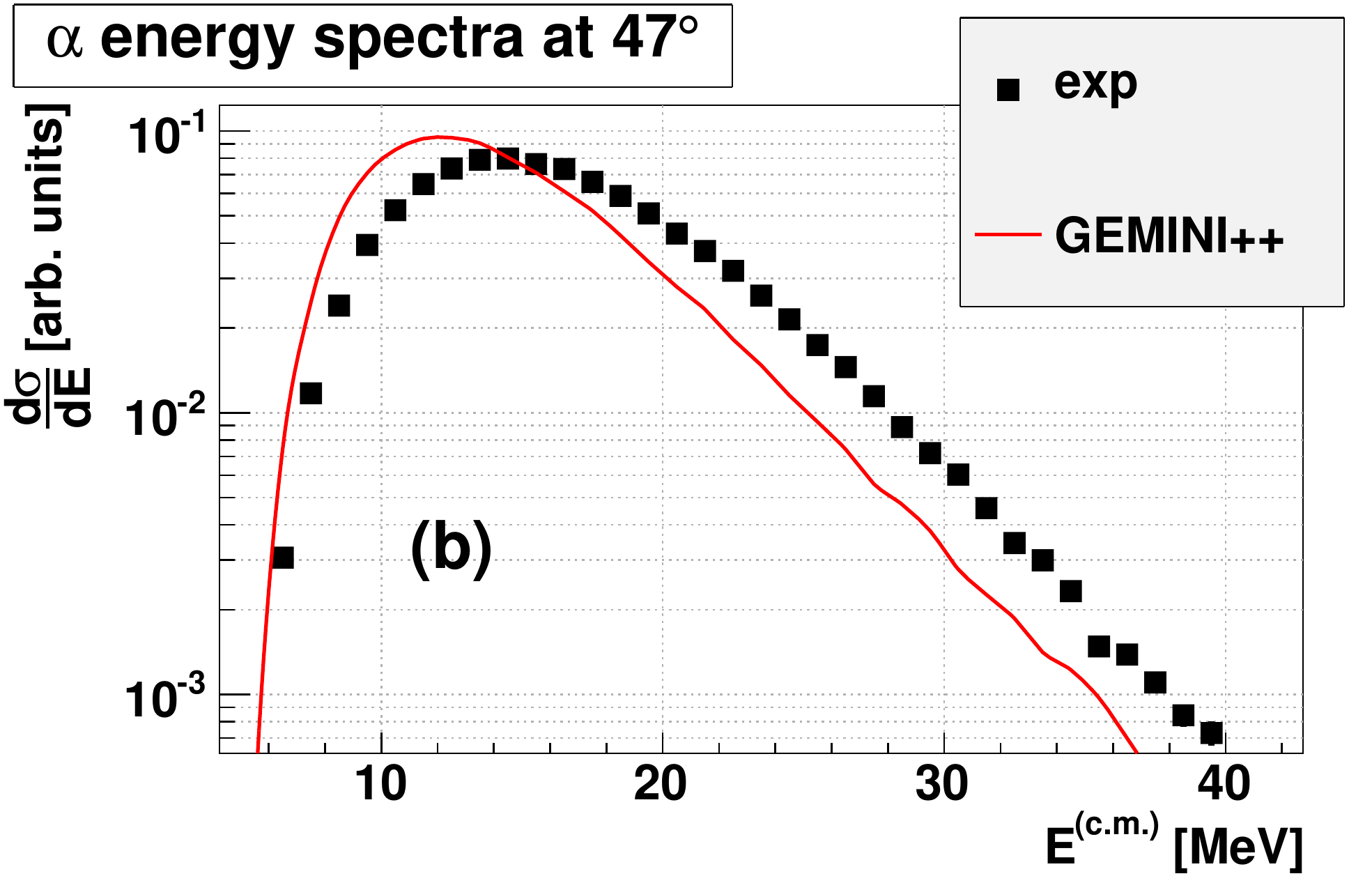}
\end{tabular}
\caption{
	(Color online) (a): center of mass energy spectra of protons in the reaction at \SI{300}{MeV} for
	experimental data (black squares)  and \gemini\ simulation (red curve) with standard parameters (from \cite{Charity10}).
	The spectra are normalized to their integrals. Data refer to particles detected in the \garf\ ring covering the polar angles
	between \ang{41} and \ang{52}. (b): same for $\alpha$-particles.
}
\label{fig:standard}
\end{figure*}

A comparison of the experimental center of mass kinetic energy spectra for light particles (black squares)
with the prediction of the \gemini\ code (red curve), run with standard values of the parameters
(those reported in \cite{Charity10}) and filtered with a software replica of the setup, is plotted
in Fig.~\ref{fig:standard} for the reaction at \SI{300}{MeV} for particles detected in the \garf\ ring covering
the polar angle between \ang{41} and \ang{52}. Side (a) concerns protons, while on side (b) $\alpha$-particle spectra are presented.
The spectra are normalized to their integrals, in order to put into evidence differences in their shapes.

From these plots a good agreement emerges between the model and experimental data in the case of protons, while
the $\alpha$-particle spectra are strongly different: in fact the experimental Coulomb barrier is higher than the simulated one.

In order to improve the model agreement in the case of $\alpha$-particles,
we investigated the effect of some model parameters, chosen among those mainly influencing the shape of the
kinetic energy spectra of evaporated particles as reported in \cite{Charity10}.

The effect of the level density parameter was investigated varying $k_0$ in the range \SIrange{6}{10}{MeV}.
According to \cite{Charity10} the standard $k_0$ value for \ce{^{88}Mo} CN is \SI{7.3}{MeV}.
Please note that according to equation (15) of \cite{Charity10}, since the $\kappa$ parameter is close to 0.1 for $A=88$, for
our system the effective level density parameter $\tilde{a}$ is almost independent of the excitation energy
and approximately equal to $\tilde{a}=\frac{A}{k_0}$ \cite{Ciemala15}.
We found that for protons the standard value $k_0$ of \cite{Charity10} is fine, while
no appreciable improvement of the agreement between simulated and experimental spectra could be obtained for
$\alpha$-particles in the investigated range. As a consequence, the standard value was kept.

In \gemini\ the effect of the source deformation due to thermally induced shape fluctuations is taken
into account by averaging the transmission coefficients over three different values calculated with three
different radius parameters of the nuclear potential
($R_0$, $R_0-\delta r$ and $R_0+\delta r$, equation (9) of reference \cite{Charity10}),
where $\delta r=w \sqrt{T}$ and T is the nuclear temperature of the daughter nucleus.
The spread of the transmission coefficient mainly influences the shape of LCP energy spectra
in the Coulomb barrier region and its effect is evident for $\alpha$ spectra, while
it is almost negligible for protons, due to their lower Coulomb barrier.
The standard \gemini\ prescription is $w=\SI{1.0}{fm}$; however, in \cite{Charity10} a case
(\ce{^{106}Cd}, \cite{Nebbia94}) is presented where $w=\SI{0}{fm}$ (corresponding to a single barrier, i.e. a spherical nucleus) fits better.
Note that the author of \cite{Charity10} doesn't exclude that this discrepancy is due to a possible contamination
from reaction mechanisms different from FE.
In any case, on the basis of this evidence, we investigated the effect of $w$ on the simulated spectra in
the range $w=\;$\SIrange{0.0}{1.5}{fm}. Proton spectra are not affected by the variation of this parameter, while for
$\alpha$-particles a slight improvement (consisting in a shift of the barrier towards the experimental value)
has been obtained by decreasing the $w$ parameter.

The other parameter able to influence the shape of energy spectra mainly for $\alpha$-particles
and mainly for light systems ($A<150$) is the shape of the yrast line; in \cite{Charity10} a shape that follows the Sierk parametrization \cite{Sierk86}
up to a threshold angular momentum and then increases linearly is proposed for light systems (as our \ce{^{88}Mo}),
obtaining very good agreement with many experimental spectra. However, it has to be noted that the explored spin
range in \cite{Charity10} (up to \SIrange{40}{50}{\planckbar}) is below the angular momenta reachable by our system
(see Table~\ref{tab:param}) and thus this recipe may not work properly in our case. Another shape of the yrast line is delivered by the RLDM of \cite{Cohen74}.

\begin{figure}[htbp]
\centering
\includegraphics[width=\columnwidth] {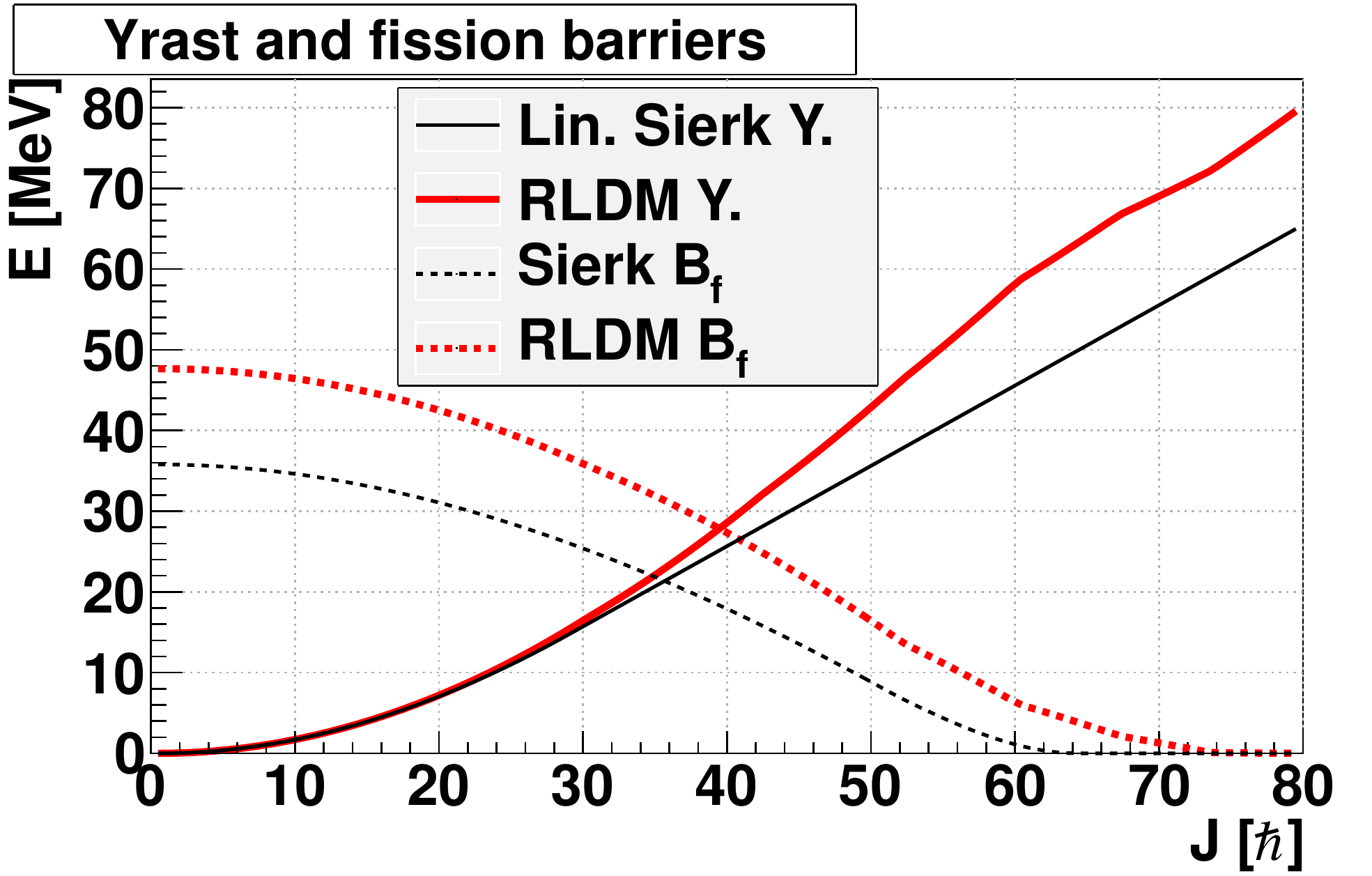}
\caption{
	(Color online) Solid curves: yrast line parametrizations for \ce{^{88}Mo} nuclei as a function
	of their angular momentum. Black thin curve: linearized Sierk yrast (prescription of \cite{Charity10}).
	Red thick curve: Rotating Liquid Drop Model yrast \cite{Cohen74}. Dotted curves: Fission barrier parametrizations for \ce{^{88}Mo}
	nuclei as a function of their angular momentum. Black thin curve: Sierk fission barrier \cite{Sierk86}.
	Red thick curve: RLDM fission barrier \cite{Cohen74}.
}
\label{fig:byrast}
\end{figure}

The two different yrast recipes (\cite{Charity10} and RLDM \cite{Cohen74})
for our system are shown in Fig.~\ref{fig:byrast} as solid curves. The two curves are very similar up to $J\sim\SI{35}{\planckbar}$;
above this value the RLDM yrast (red thick curve) tends to be similar to that of a spherical nucleus (not shown).
In the same Fig.~\ref{fig:byrast} we compare also two recipes for the fission barrier,
 the Sierk prescription \cite{Sierk86} (black thin dotted curve) and the RLDM \cite{Cohen74}
(red thick dotted curve). At a fixed angular momentum the RLDM barrier is higher than the Sierk one,
thus reducing the fission probability for the ER. This has influence not only on the shape of the particle energy
spectra but also on their multiplicities.

\begin{figure}[htpb]
\centering
\includegraphics[width=\columnwidth] {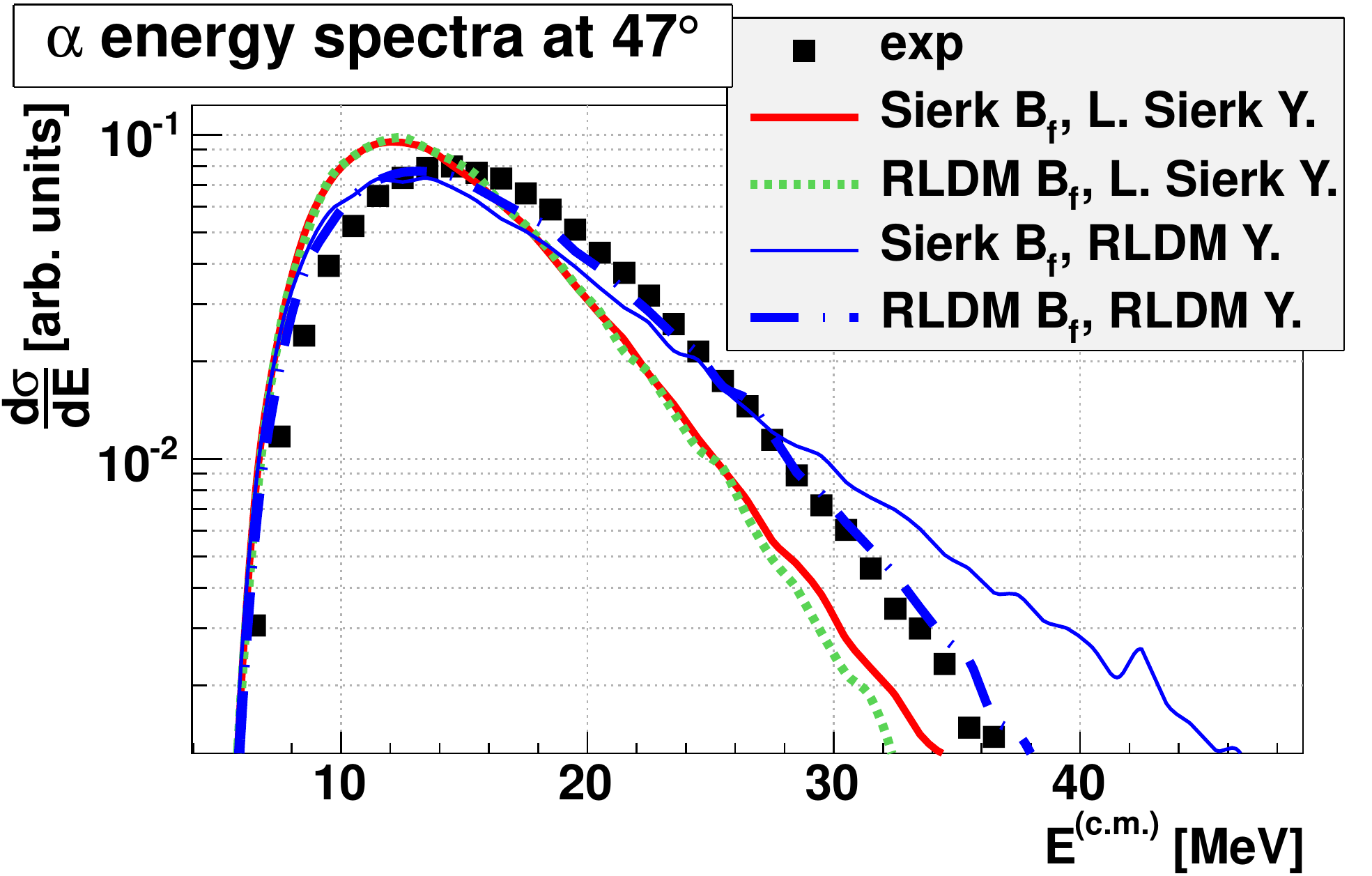}
\caption{
	(Color online) Center of mass energy spectra of $\alpha$-particles in the reaction at
	\SI{300}{MeV} for experimental data (black squares) and \gemini\ simulations with different prescriptions for
	the fission barrier and the yrast curve. The standard value (Sierk fission barrier and Linearized Sierk yrast)
	is the red continuous thick curve; the green dotted curve corresponds to standard yrast (Linearized Sierk yrast)
	and RLDM fission barrier; the thin blue curve corresponds to standard fission barrier (Sierk fission barrier) and RLDM yrast;
	the blue dash-dotted curve corresponds to RLDM fission barrier and RLDM yrast.
	Spectra are normalized to their integrals. Data refer to particles detected in the \garf\ ring covering
	the polar angles between \ang{41} and \ang{52}.
}
\label{fig:yrastbf}
\end{figure}

Proton spectra are almost independent of the adopted parametrization, while for $\alpha$-particles the best
results are obtained when the RLDM barrier is coupled to the RLDM yrast, as shown in Fig.~\ref{fig:yrastbf}
(blue thick dash-dotted curve).
We mention that this particular \gemini\ parameter set corresponds to the choice made in \cite{Ciemala15}.

\begin{figure}[htpb]
\centering
\includegraphics[width=\columnwidth] {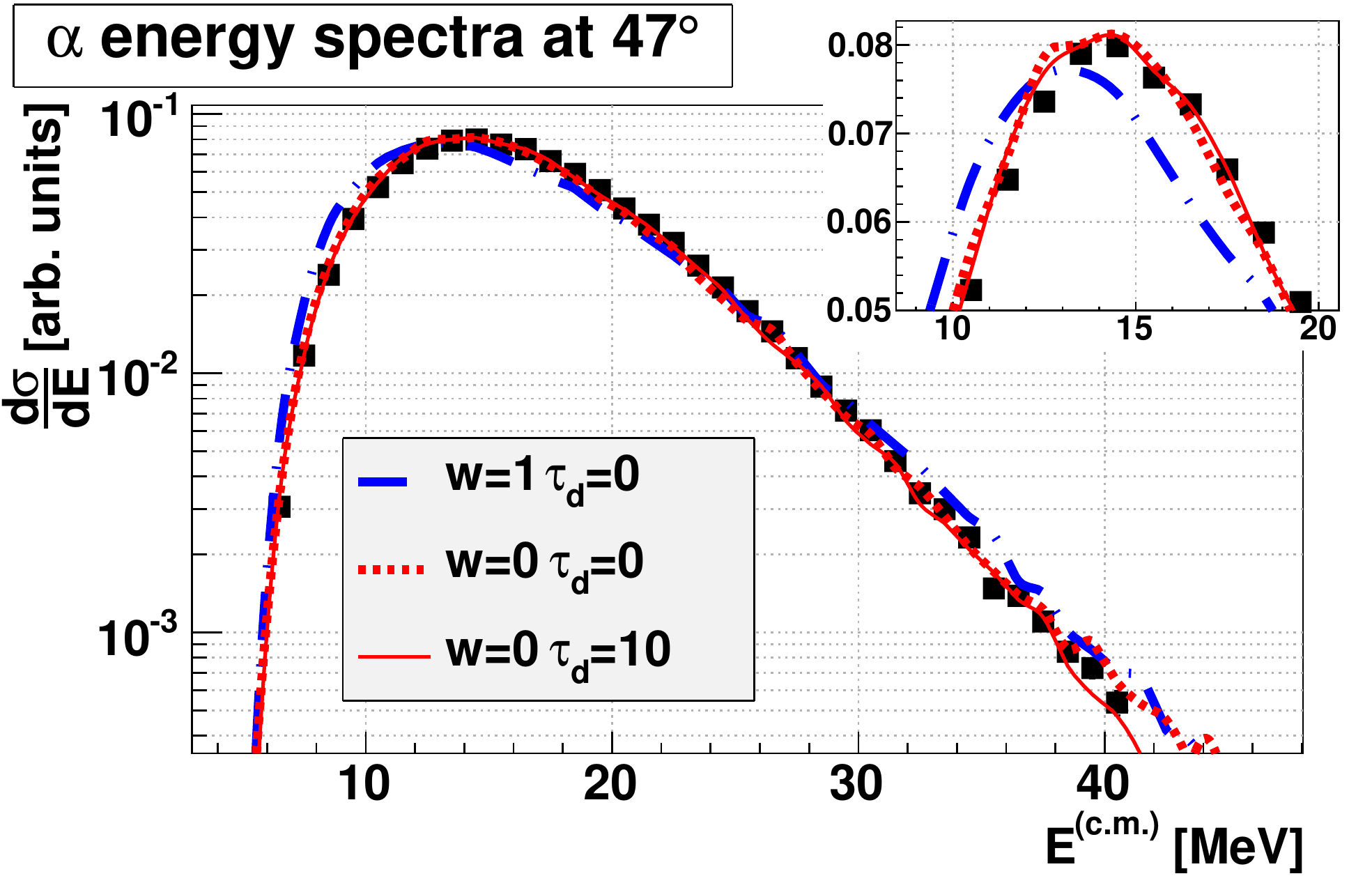}
\caption{
	(Color online) Center of mass energy spectra of $\alpha$-particles in the reaction at \SI{300}{MeV} for
	experimental data (black squares) and \gemini\ simulations with different prescriptions.
	Blue curve: RLDM fission barrier, RLDM yrast line, $w=\SI{1.0}{fm}$ and $\tau_\mathrm{d}=\SI{0}{zs}$.
	Red curve: RLDM fission barrier, RLDM yrast line, $w=\SI{0}{fm}$ and $\tau_\mathrm{d}=\SI{0}{zs}$.
	Red dotted curve: RLDM fission barrier, RLDM yrast line, $w=\SI{0}{fm}$ and $\tau_\mathrm{d}=\SI{10}{zs}$.
	In the inset the barrier region is zoomed with a linear ordinate scale. Spectra are normalized to their integrals.
	Data refer to particles detected in the \garf\ ring covering the polar angles between \ang{41} and \ang{52}.
}
\label{fig:yrastbfw}
\end{figure}

The agreement for the energy spectra of $\alpha$-particles can be further improved if we couple the RLDM barrier
and RLDM yrast prescription with the choice of $w=\SI{0}{fm}$, i.e. if we switch off thermal barrier fluctuations,
moving towards a spherical nucleus, as it is shown in Fig.~\ref{fig:yrastbfw}, blue curve; in particular,
the better agreement obtained in the barrier zone (zoomed in the inset) has to be noted.

\begin{figure*}[htpb]
\centering
\begin{tabular}{ccc}
\includegraphics[width=0.3\textwidth] {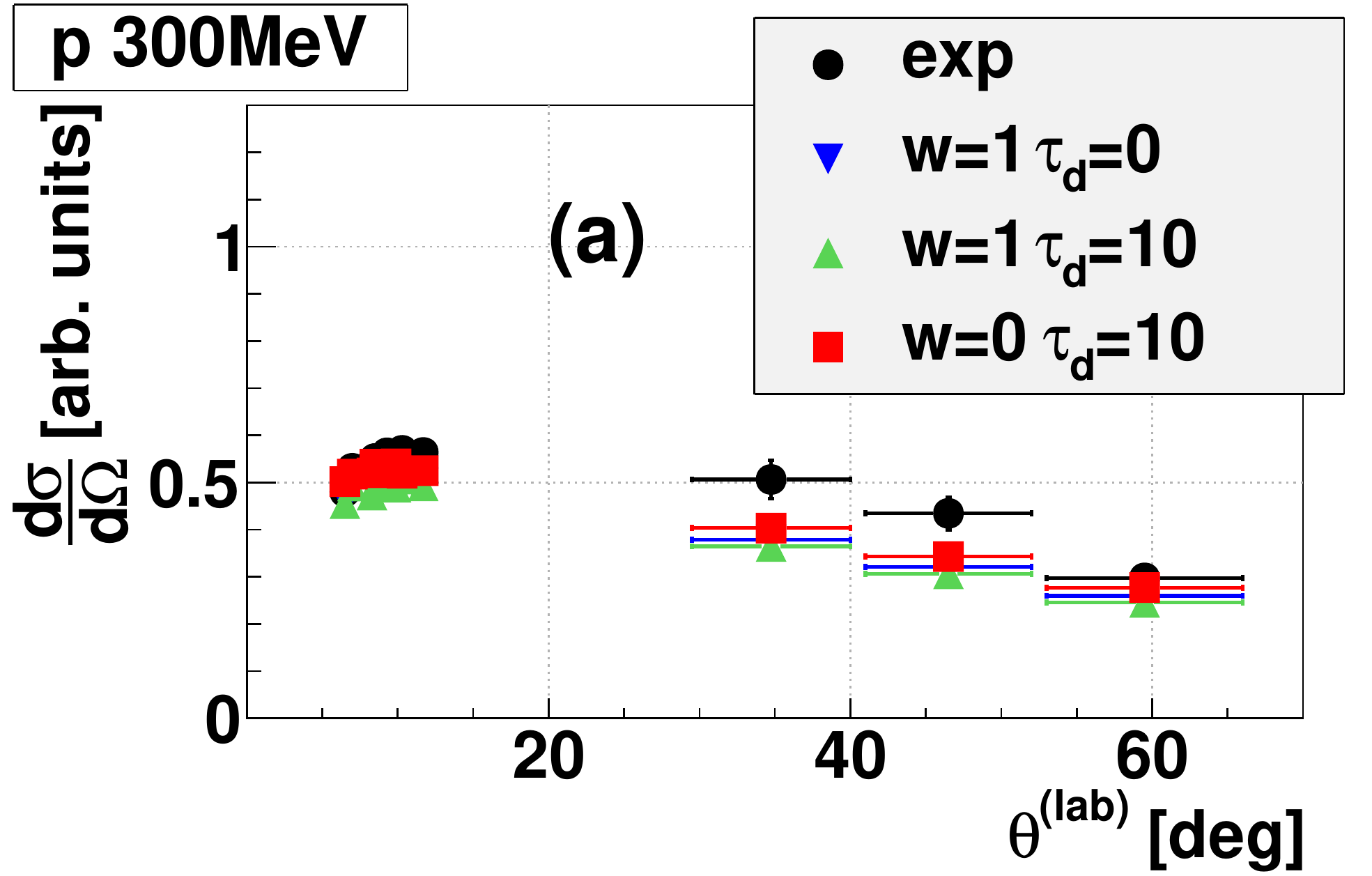} &\includegraphics[width=0.3\textwidth] {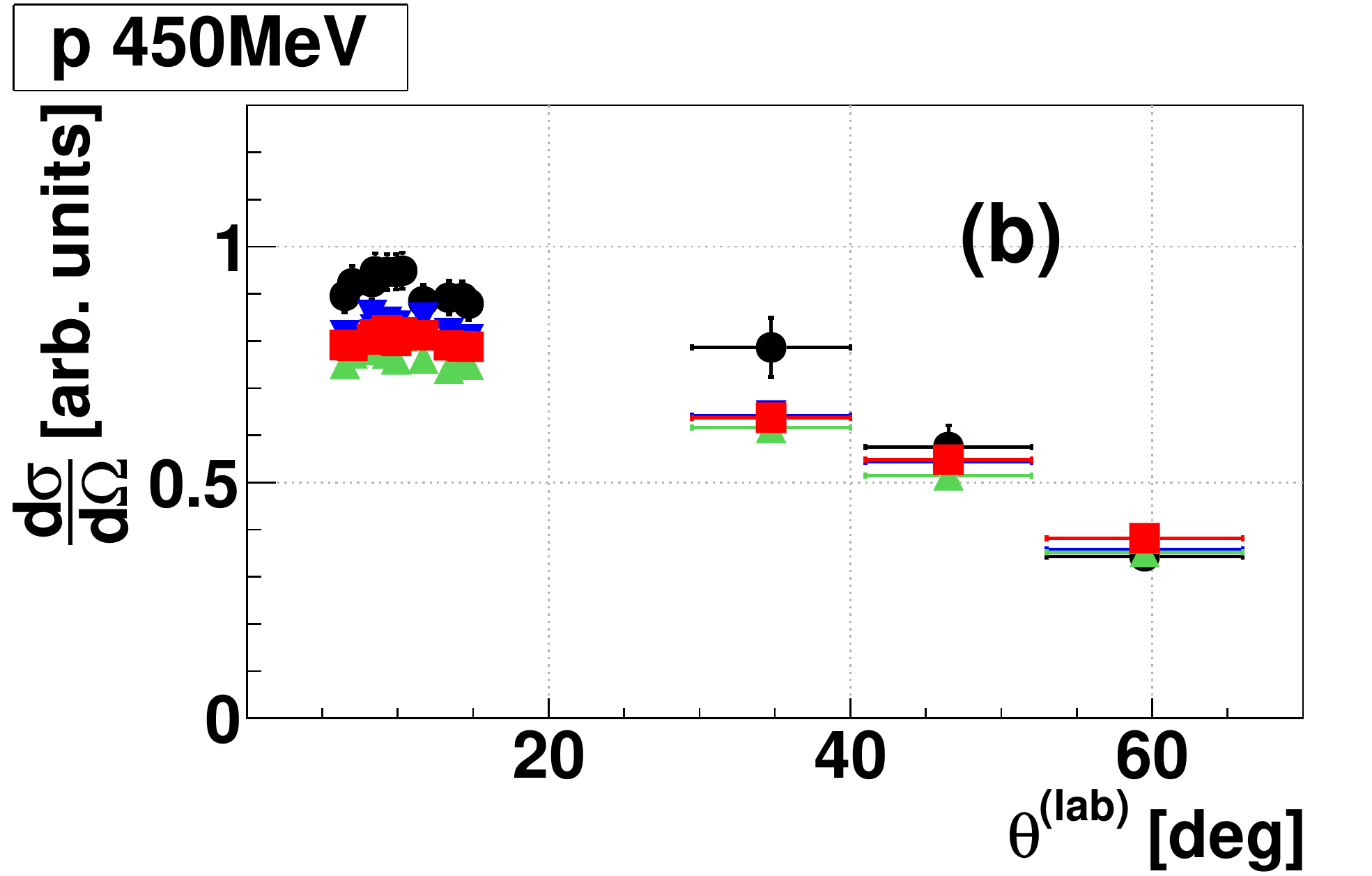} &%
\includegraphics[width=0.3\textwidth] {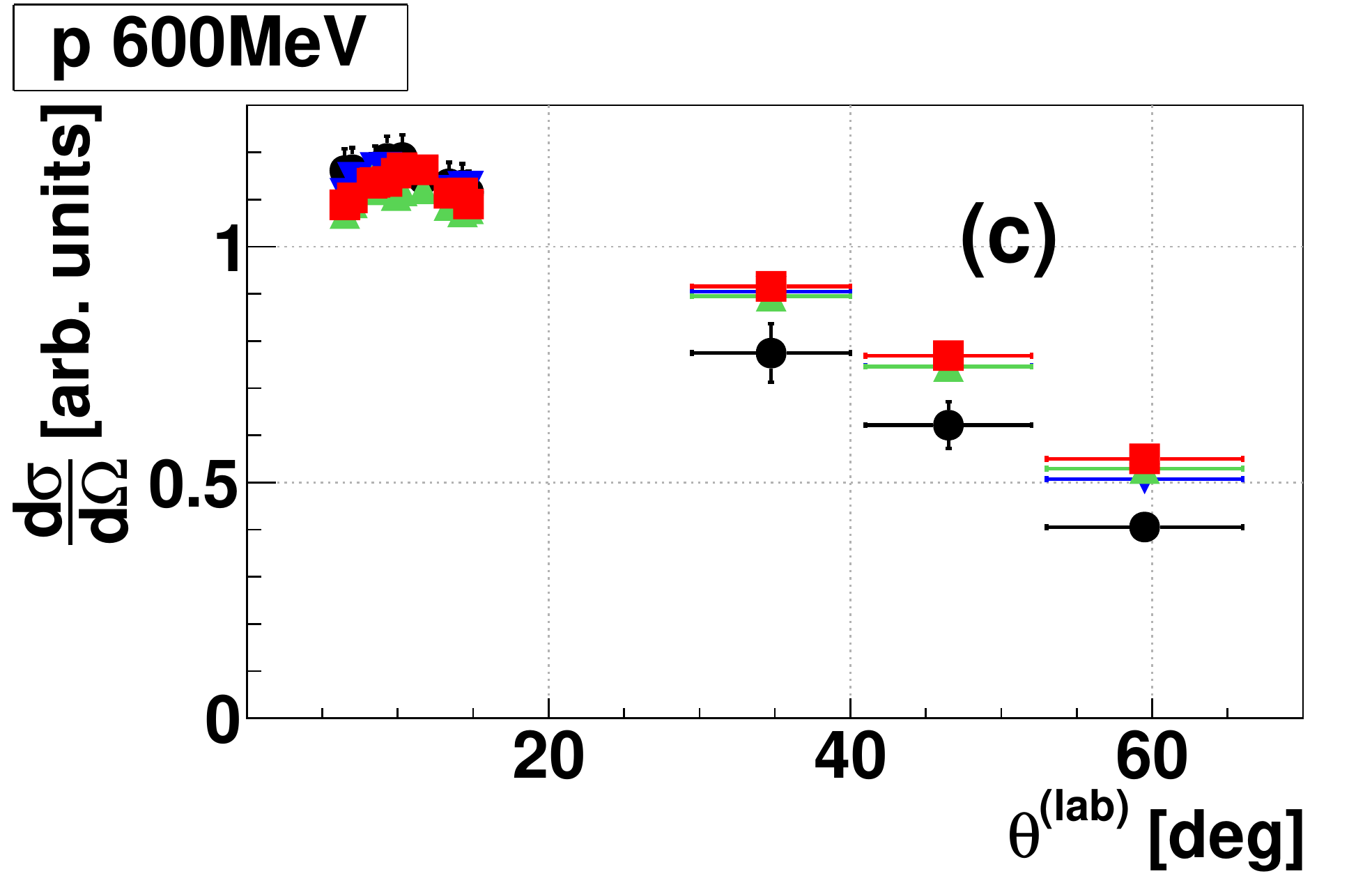}\\
\includegraphics[width=0.3\textwidth] {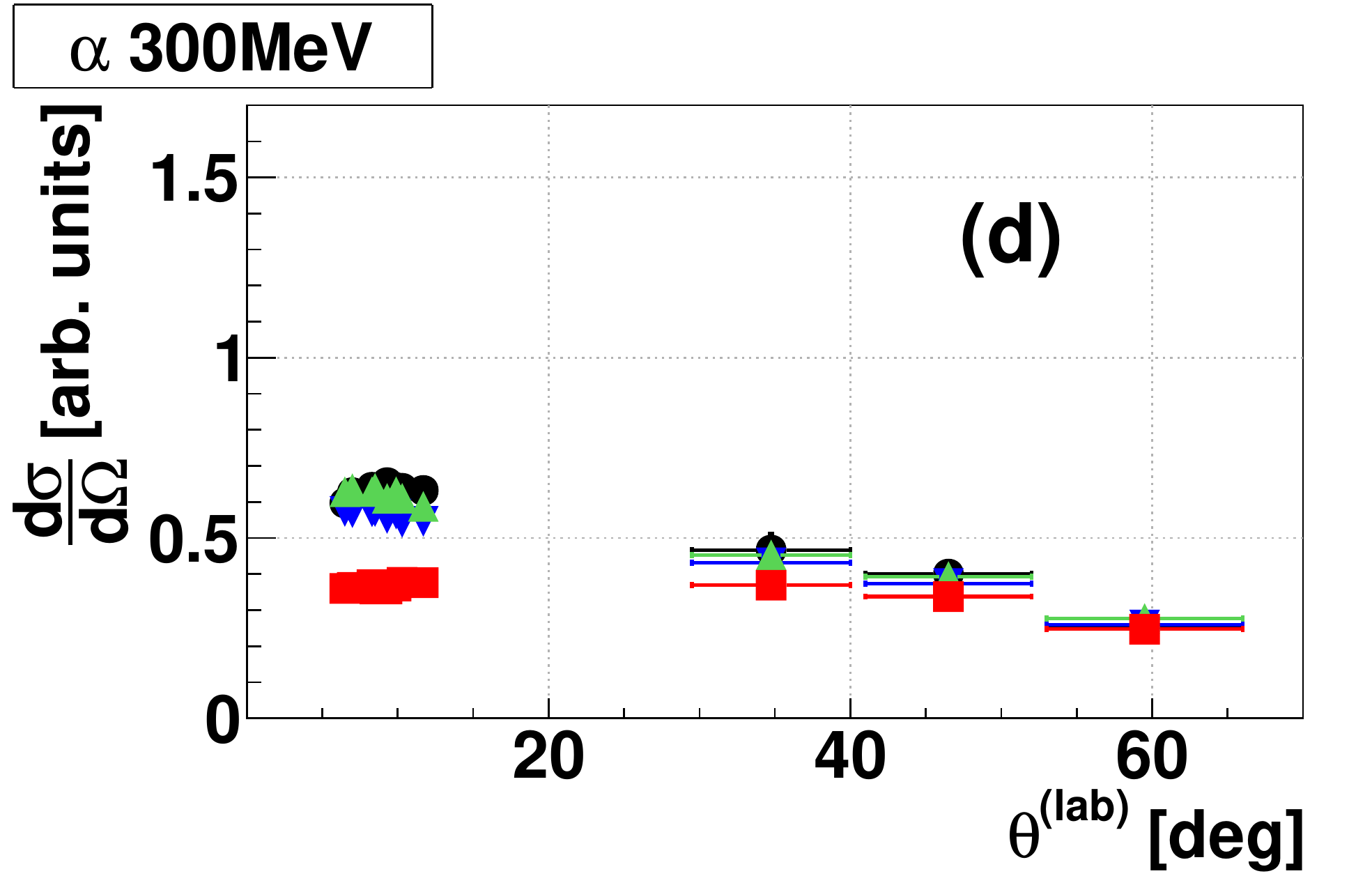} &\includegraphics[width=0.3\textwidth] {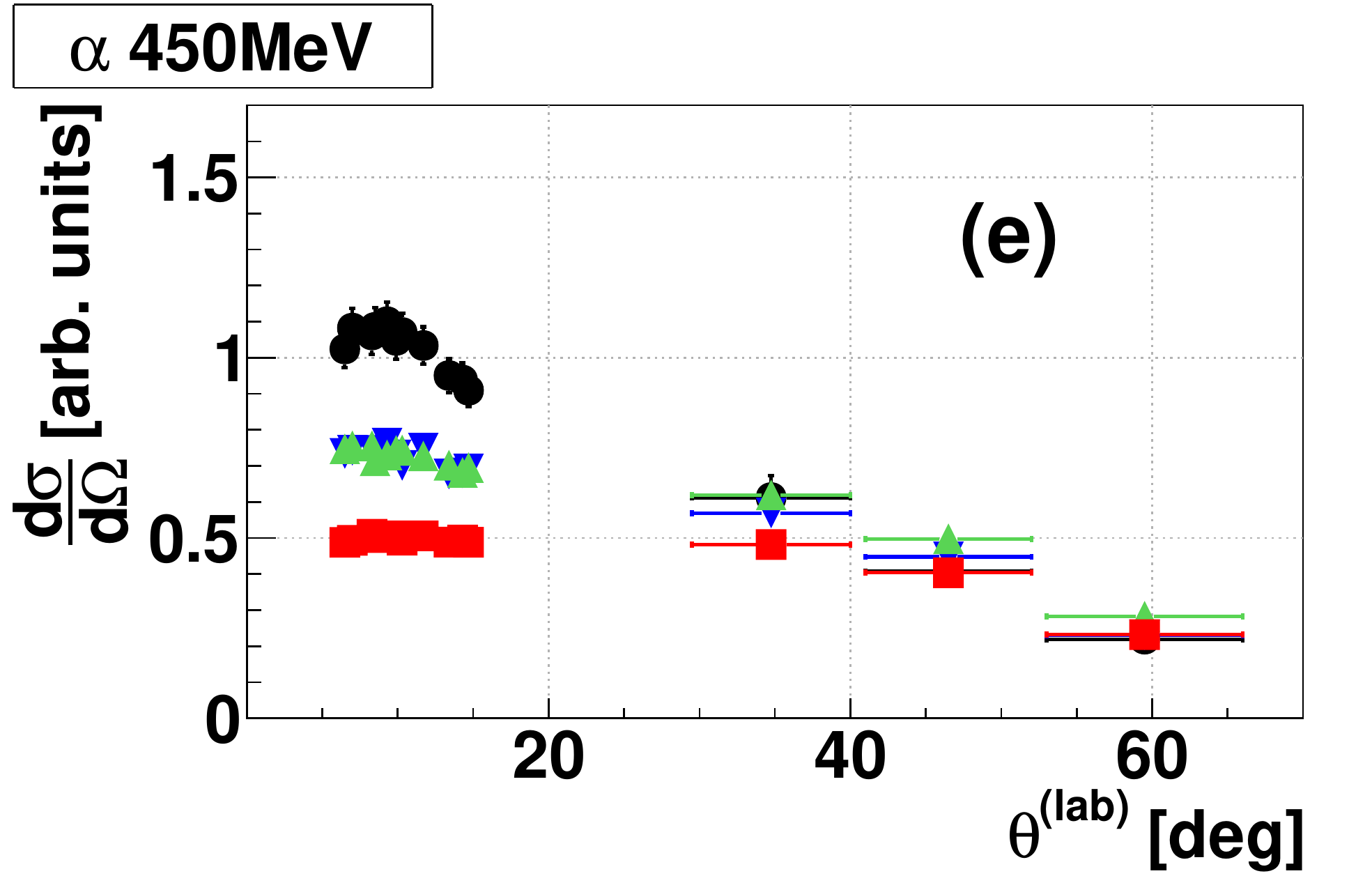} &%
\includegraphics[width=0.3\textwidth] {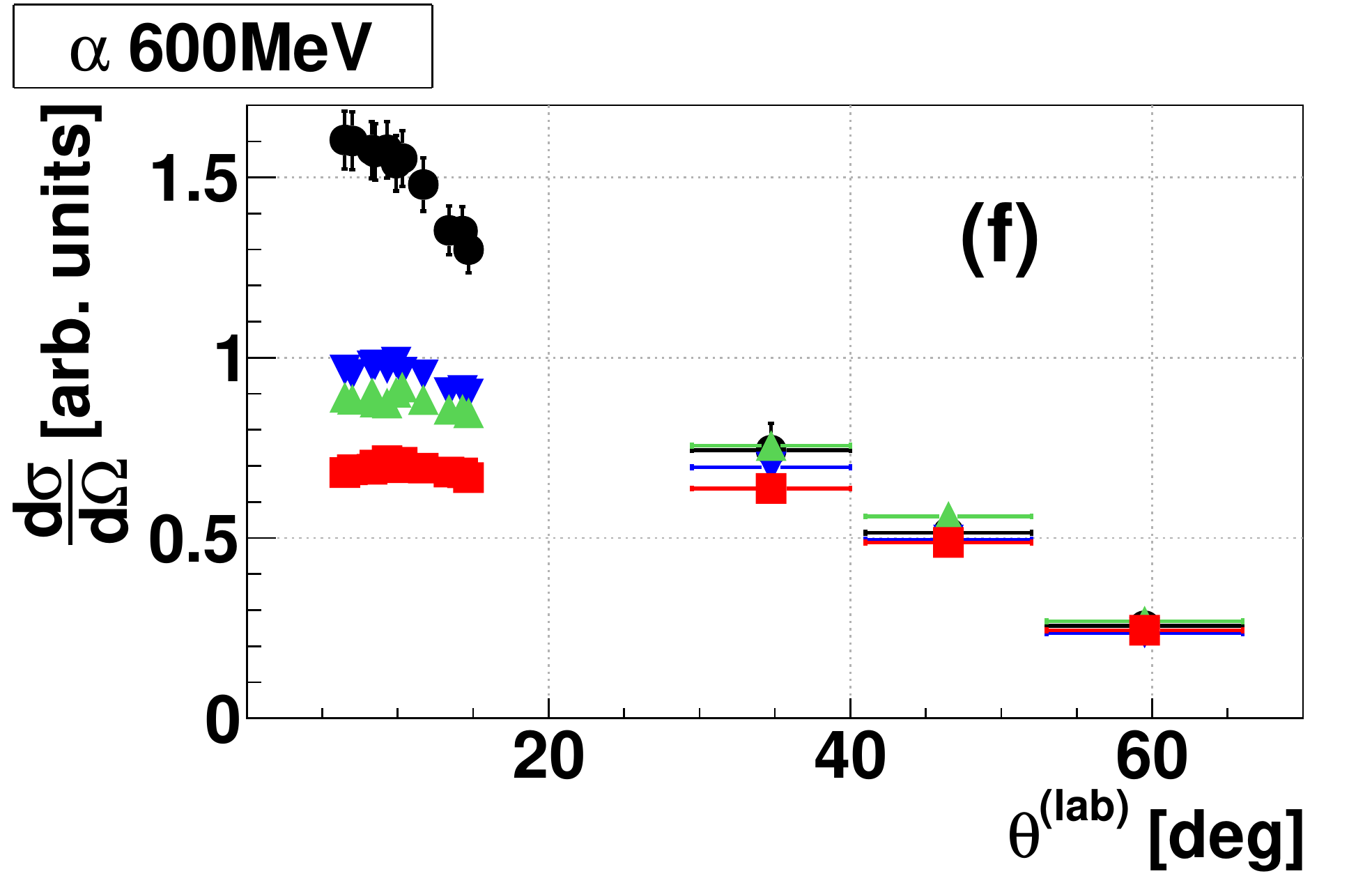}\\
\end{tabular}
\caption{
	(Color online) Differential cross section as a function of the lab polar angle.
	Panels (a), (b), (c): protons. Panels (d), (e), (f): $\alpha$-particles. (a) and (d): \SI{300}{MeV};
	(b) and (e): \SI{450}{MeV}; (c) and (f): \SI{600}{MeV}. Full black circles: experimental data.
	Red squares: \gemini\ with RLDM yrast line, RLDM fission barrier, $w=\SI{0}{fm}$, $\tau_\mathrm{d}=\SI{10}{zs}$
	(best choice for energy spectra). Blue down triangles: \gemini\ with RLDM yrast line, RLDM fission barrier,
	$w=\SI{1.0}{fm}$, $\tau_\mathrm{d}=\SI{0}{zs}$. Green up triangles: \gemini\ with RLDM yrast line, RLDM fission barrier,
	$w=\SI{1.0}{fm}$, $\tau_\mathrm{d}=\SI{10}{zs}$.
}
\label{fig:angoli}
\end{figure*}

The time delay for fission $\tau_\mathrm{d}$ has no influence on the particle energy spectra
(Fig.~\ref{fig:yrastbfw}, comparison between red continuous and red dotted curve), while it affects the fission probability
and the angular distribution of the emitted particles (as it will be discussed in Fig.~\ref{fig:angoli}).

\begin{figure*}[htpb]
\centering
\begin{tabular}{ccc}
\includegraphics[width=0.3\textwidth] {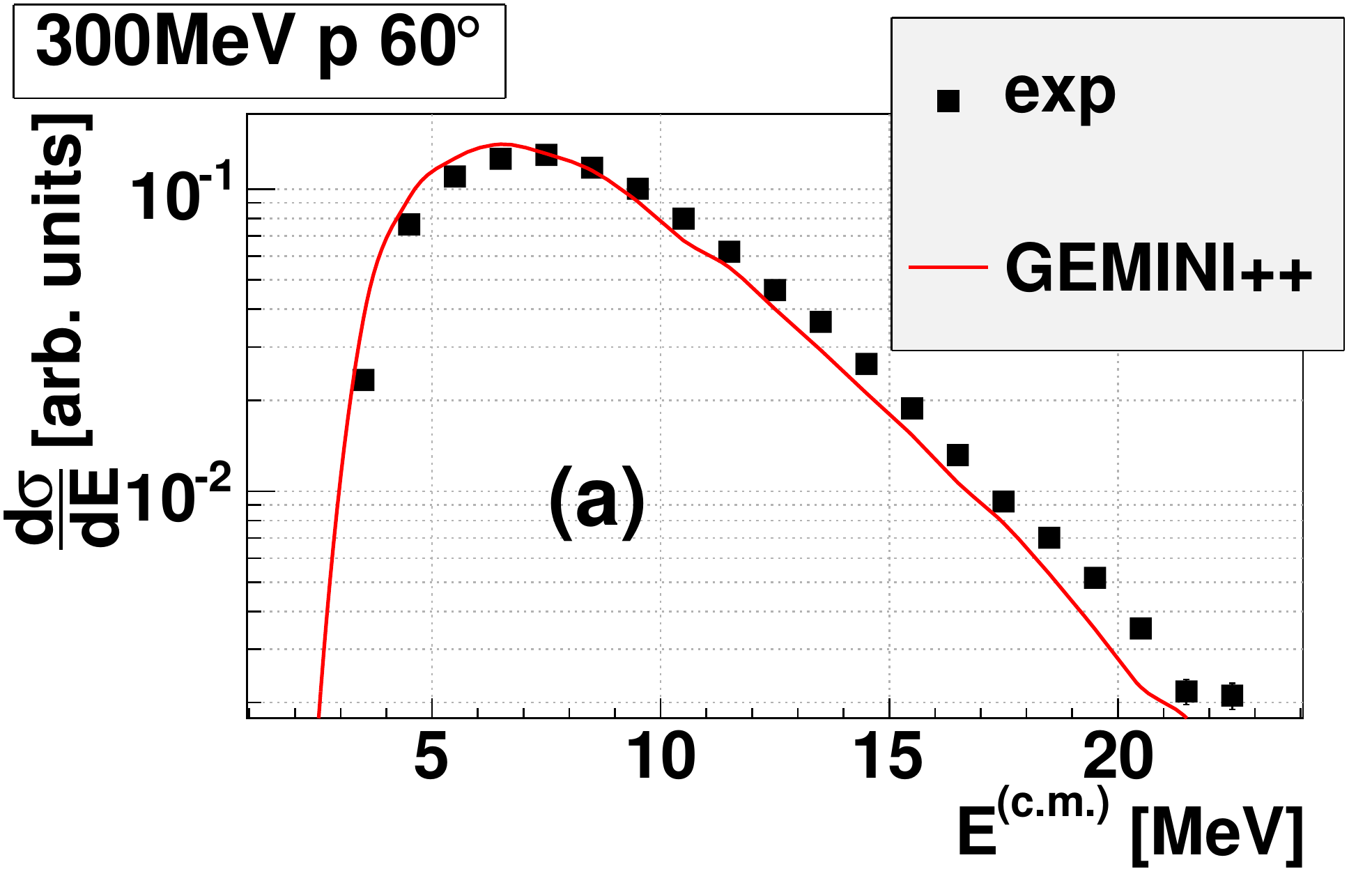} & \includegraphics[width=0.3\textwidth] {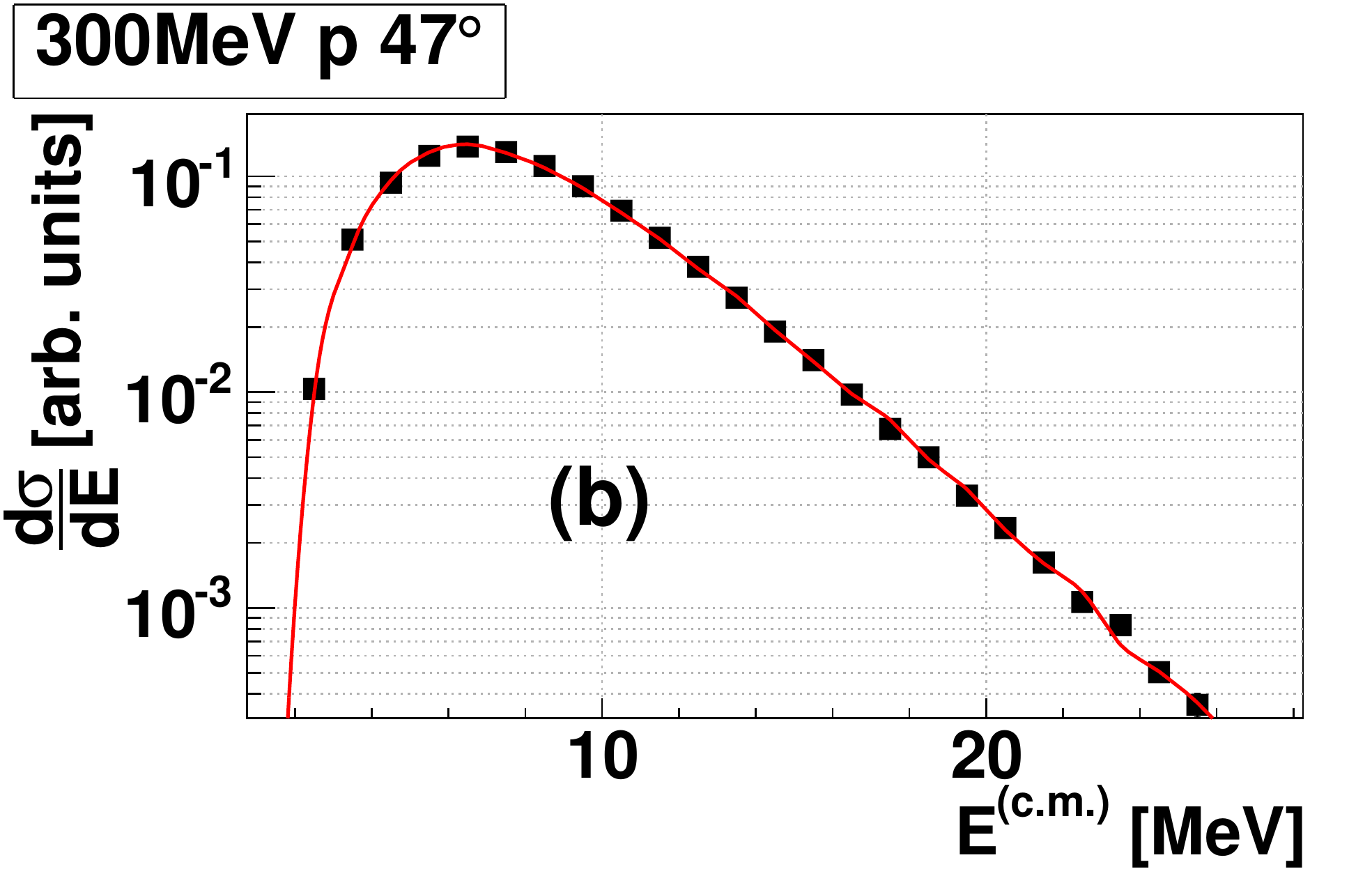} &%
\includegraphics[width=0.3\textwidth] {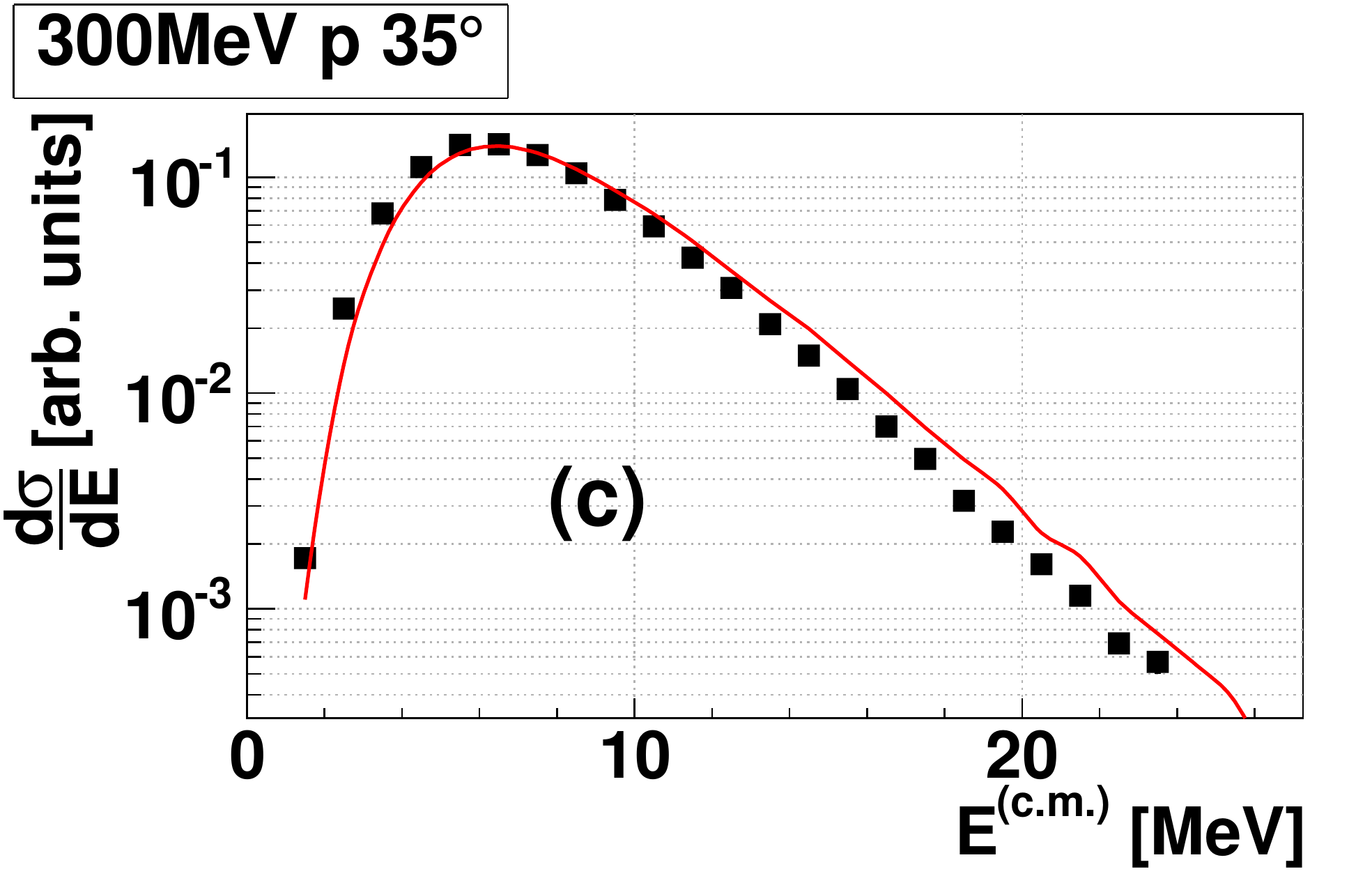}\\
\includegraphics[width=0.3\textwidth] {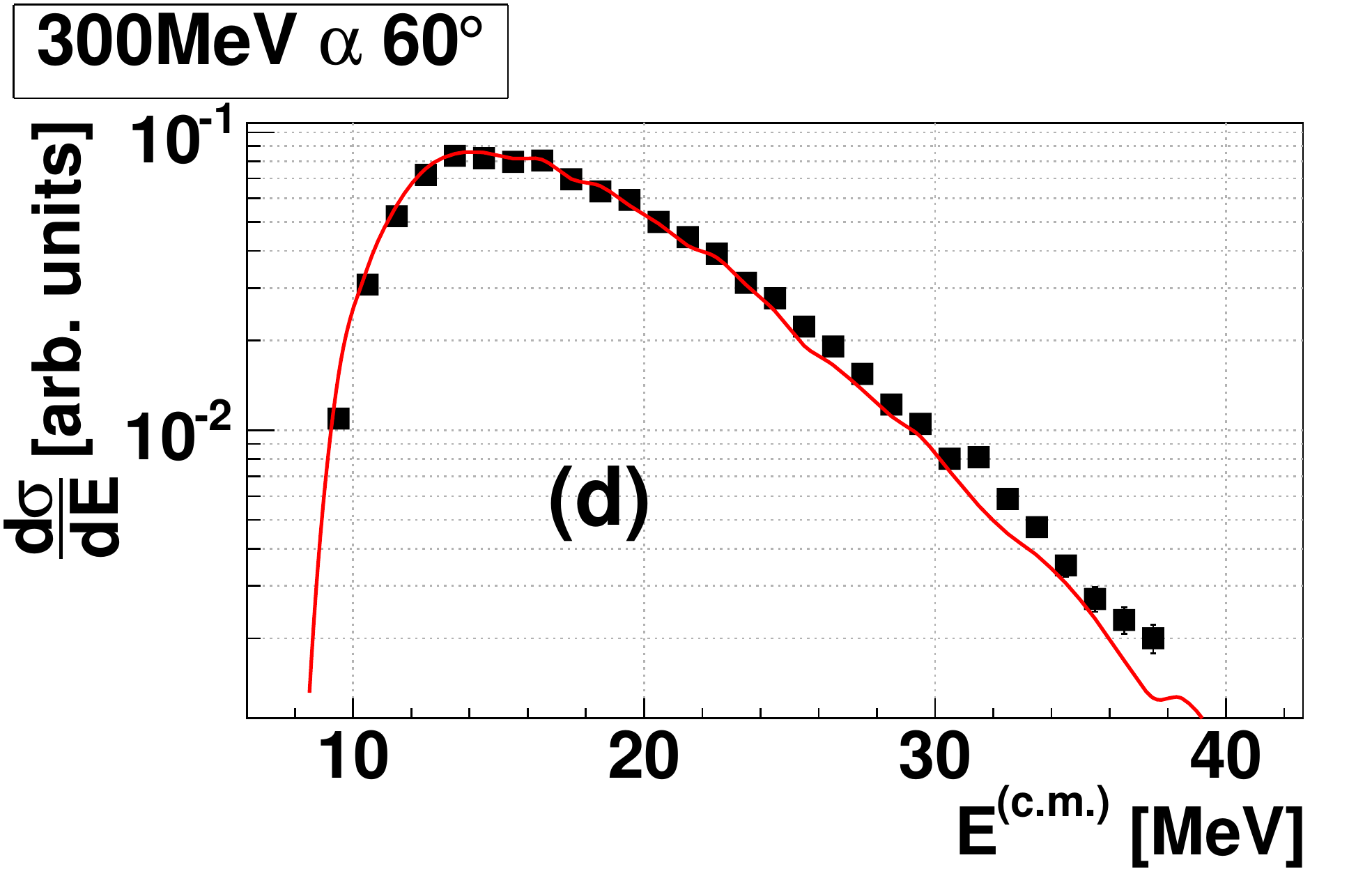} & \includegraphics[width=0.3\textwidth] {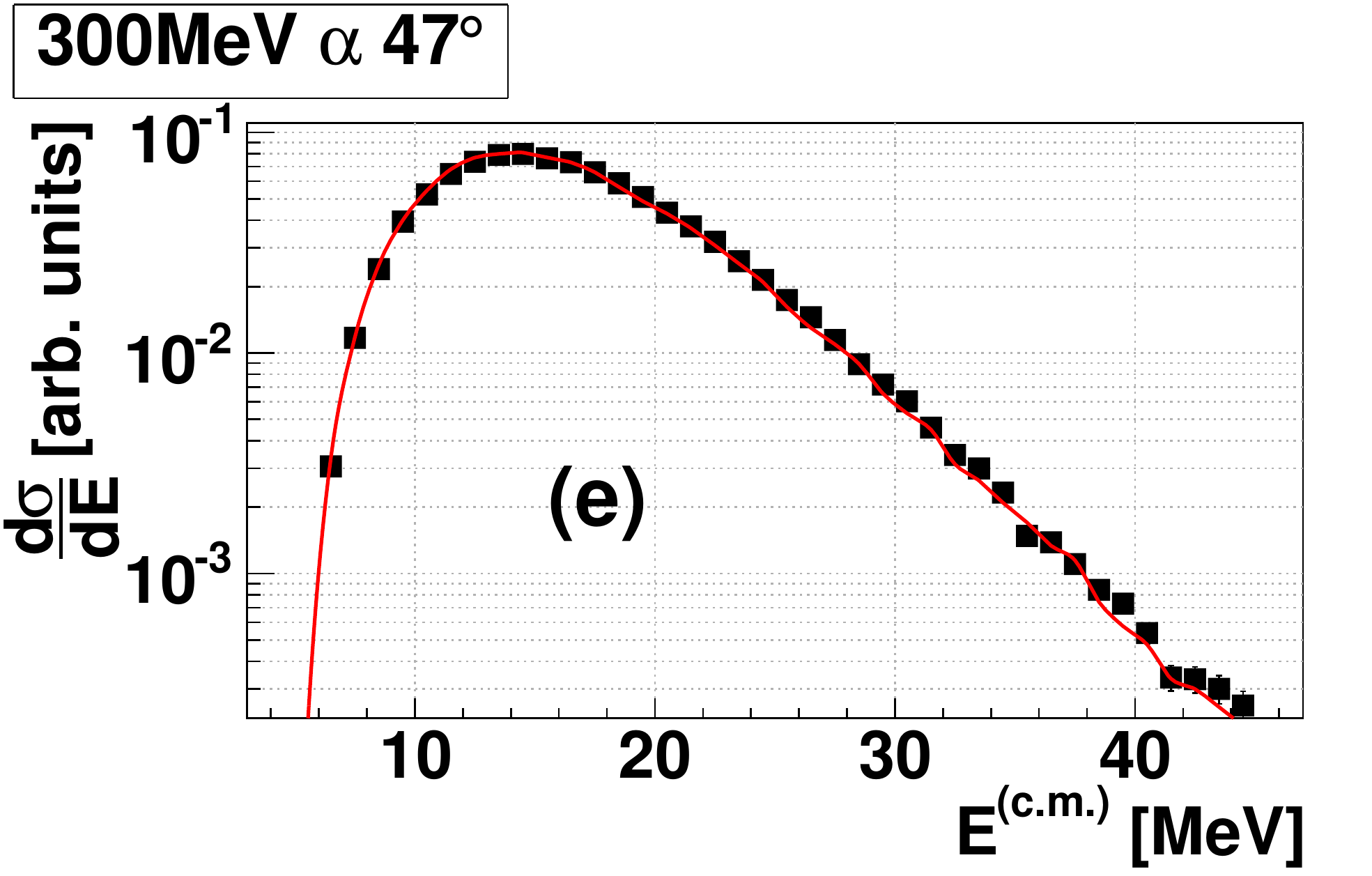} &%
\includegraphics[width=0.3\textwidth] {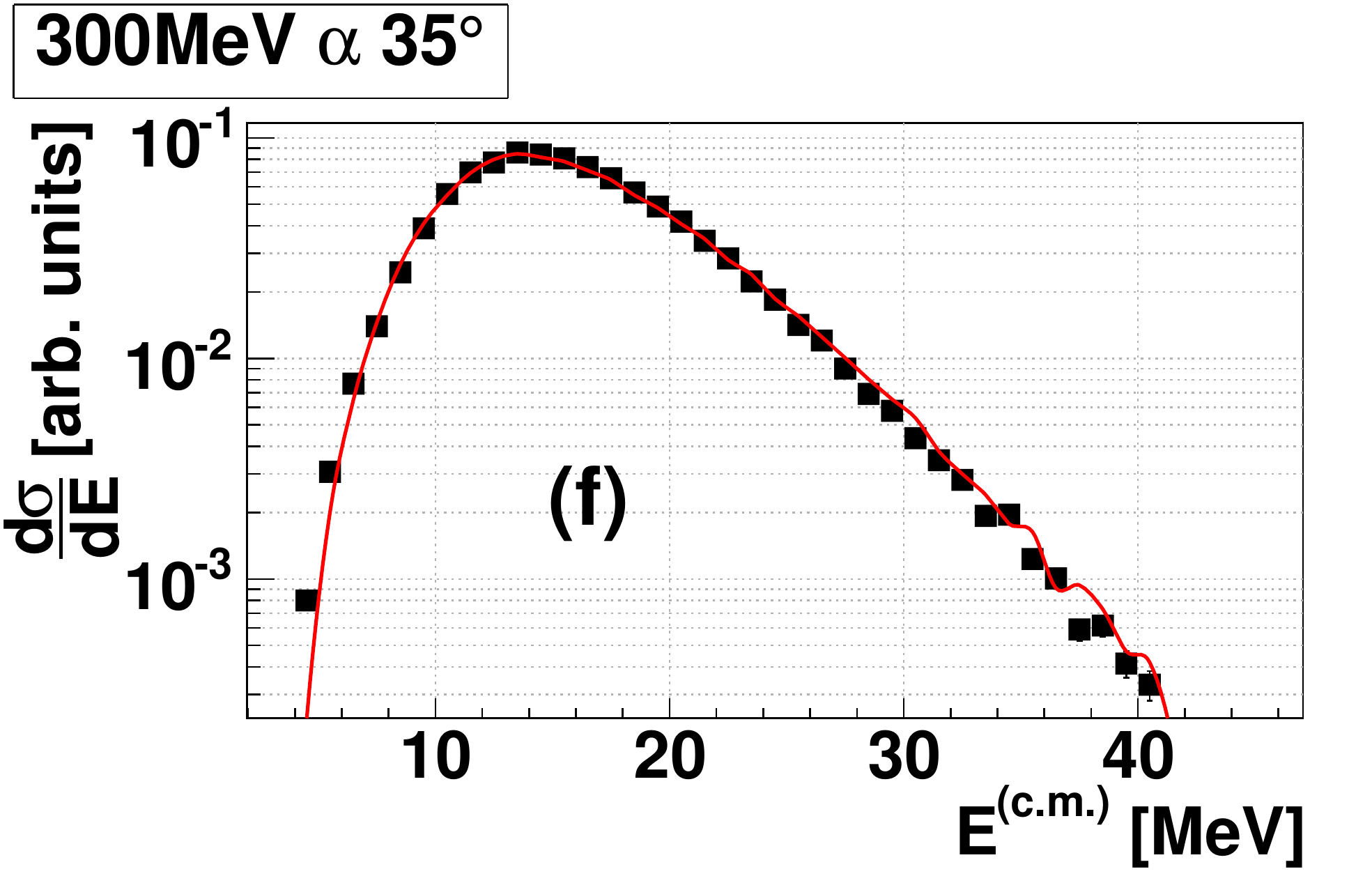}\\
\end{tabular}
\caption{
	(Color online) Center of mass energy spectra for LCPs detected at different lab angles for the reaction at \SI{300}{MeV}.
	Experimental data correspond to black squares. Simulated data have been obtained with $k_0=\SI{7.3}{MeV}$,
	RLDM fission barrier, RLDM yrast line, $w=\SI{0}{fm}$, $\tau_\mathrm{d}=\SI{10}{zs}$. Spectra are normalized to their integral.
	(a), (d): $\ang{53.0}\leq\vartheta_\mathrm{lab}\leq\ang{66.0}$; (b), (e): $\ang{41.0}\leq\vartheta_\mathrm{lab}\leq\ang{52.0}$;
	(c), (f): $\ang{29.5}\leq\vartheta_\mathrm{lab}\leq\ang{40.0}$. Panels (a), (b), (c) correspond to protons, while
	panels (d), (e), (f) correspond to $\alpha$-particles.
}
\label{fig:ring300}
\end{figure*}

The experimental center of mass energy spectra of protons (panels (a), (b), (c)) and $\alpha$-particles (panels (d), (e), (f)) for
three angular rings of \garf\ are compared in Fig.~\ref{fig:ring300} with the corresponding simulated distributions obtained
with the ``best'' set of parameters so far discussed
($k_0=\SI{7.3}{MeV}$, RLDM fission barrier, RLDM yrast line, $w=\SI{0}{fm}$, $\tau_\mathrm{d}=\SI{10}{zs}$).
The agreement between experimental data and simulation is reasonable at all angles accessible with \garf .

Calculations have been extended to the two higher beam energies assuming the parameter set tuned for the \SI{300}{MeV} reaction.

\begin{figure*}[htpb]
\centering
\begin{tabular}{ccc}
\includegraphics[width=0.3\textwidth] {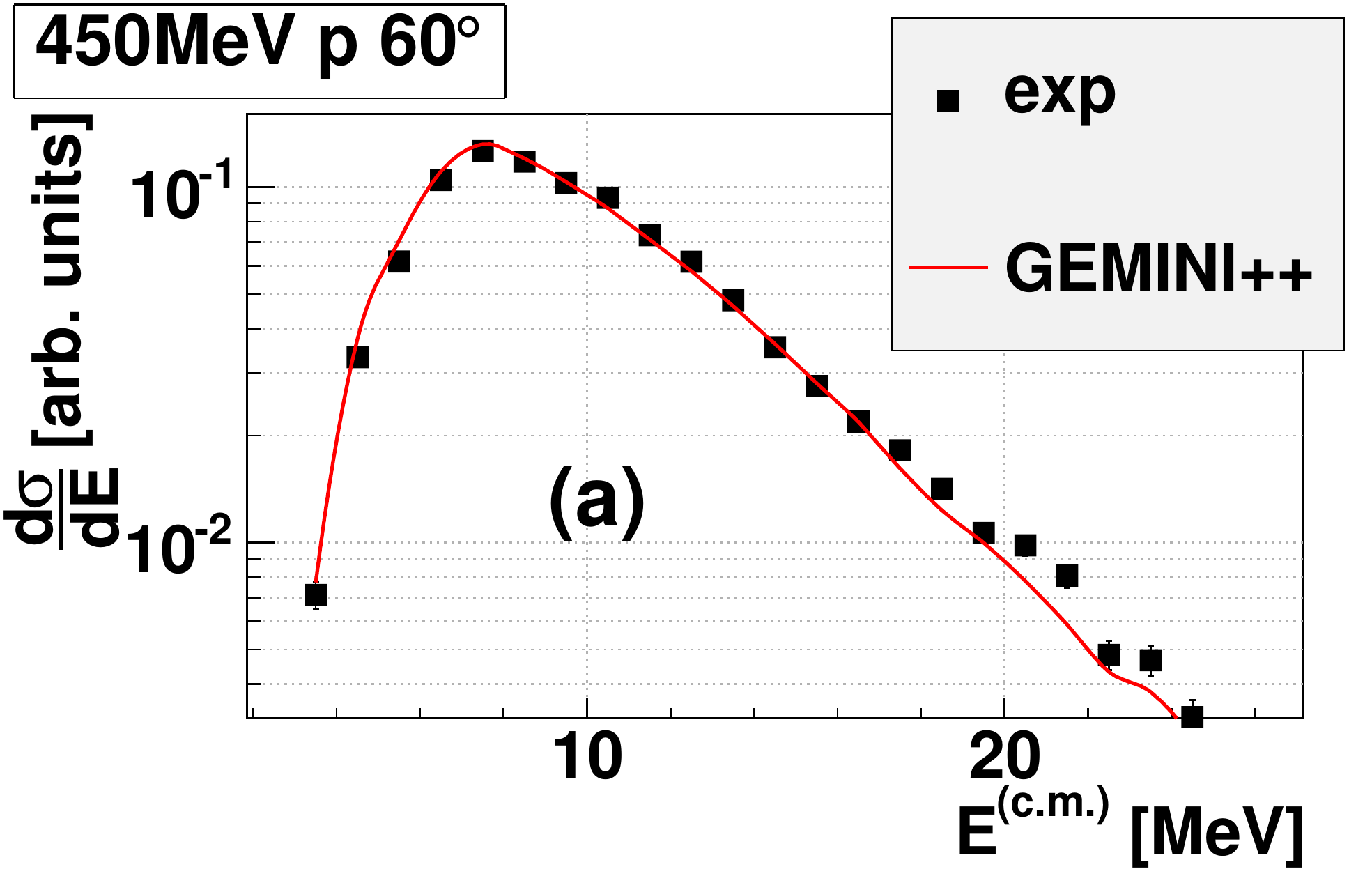} & \includegraphics[width=0.3\textwidth] {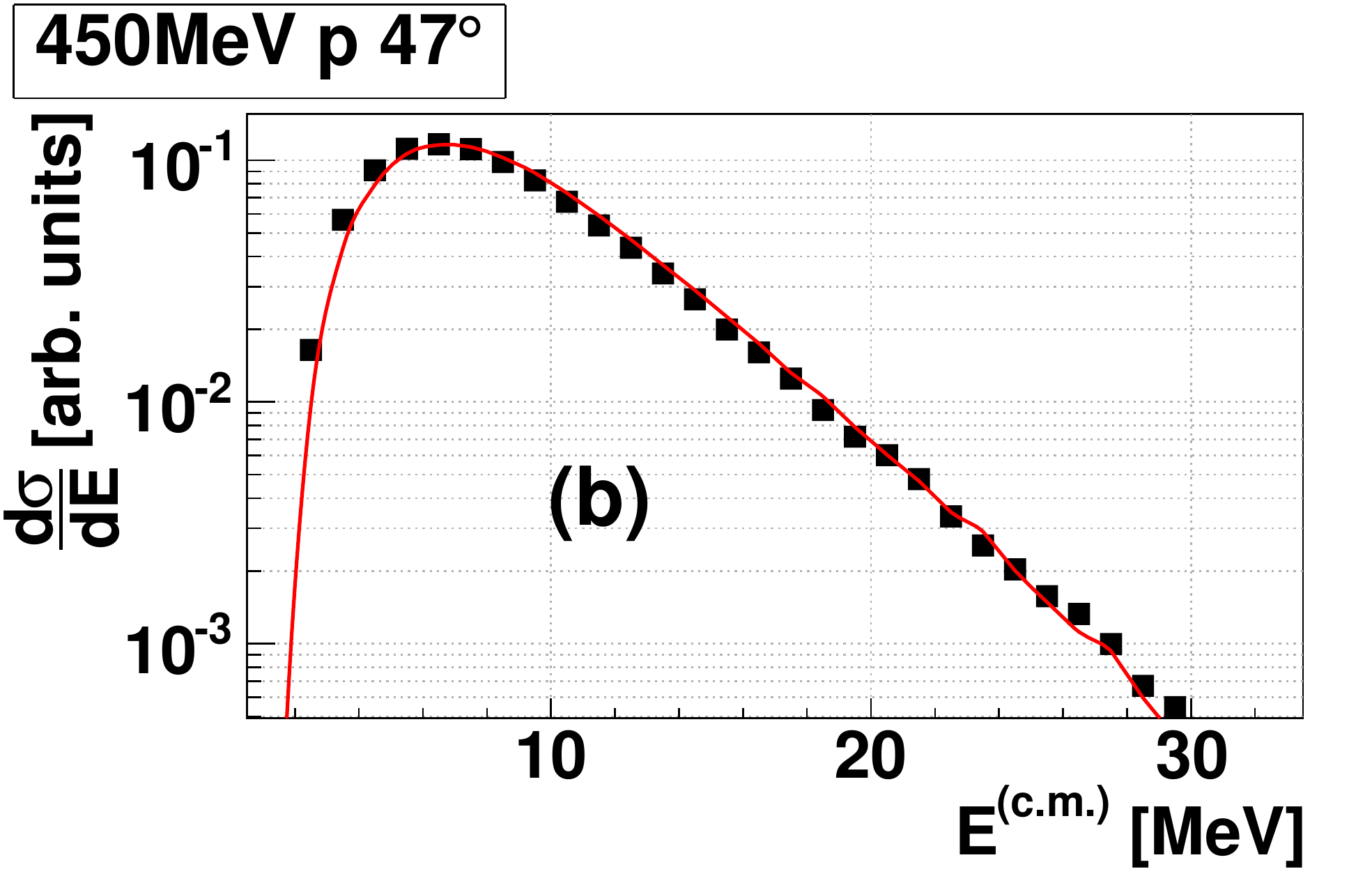} &%
\includegraphics[width=0.3\textwidth] {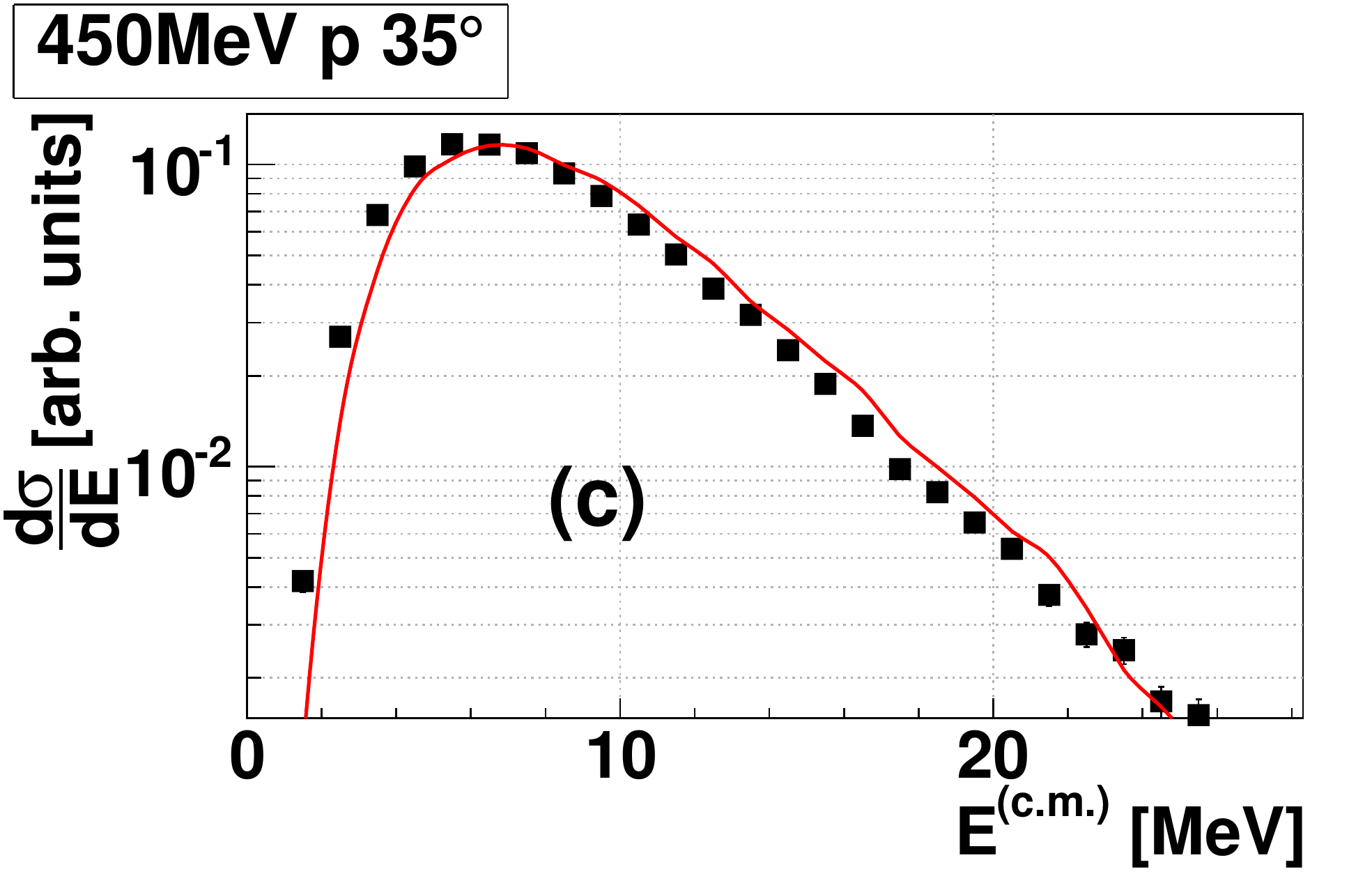}\\
\includegraphics[width=0.3\textwidth] {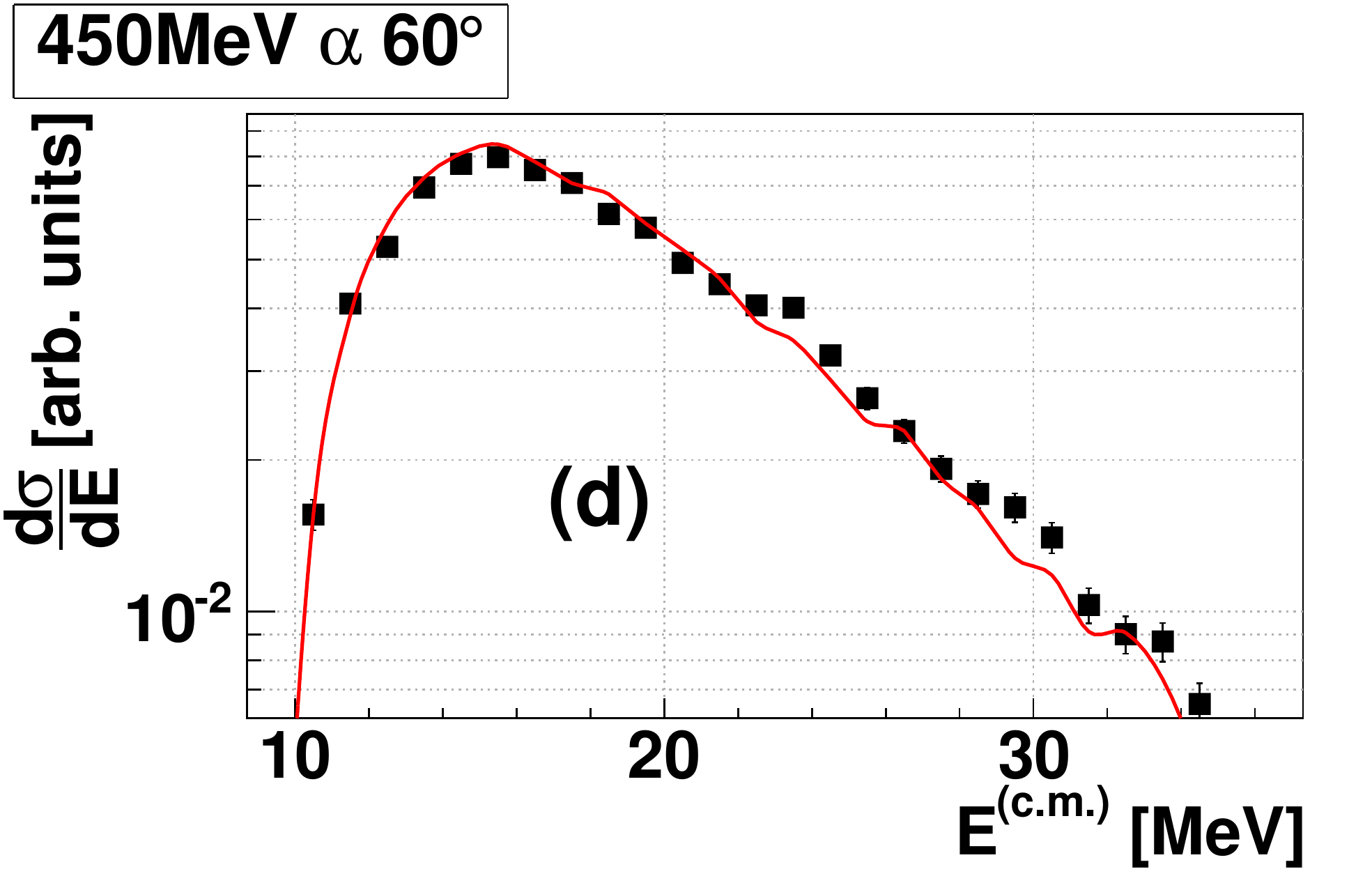} & \includegraphics[width=0.3\textwidth] {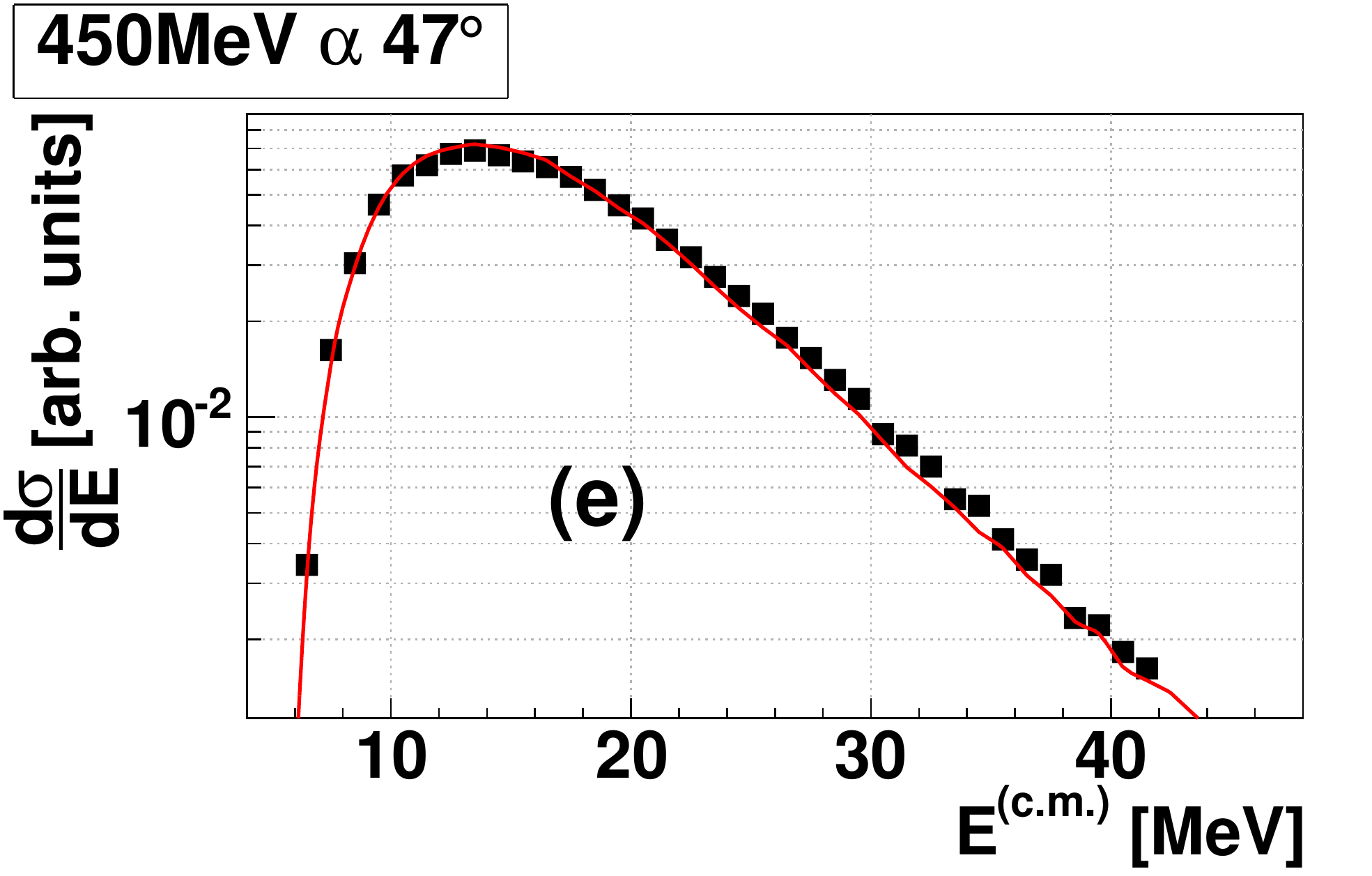} &%
\includegraphics[width=0.3\textwidth] {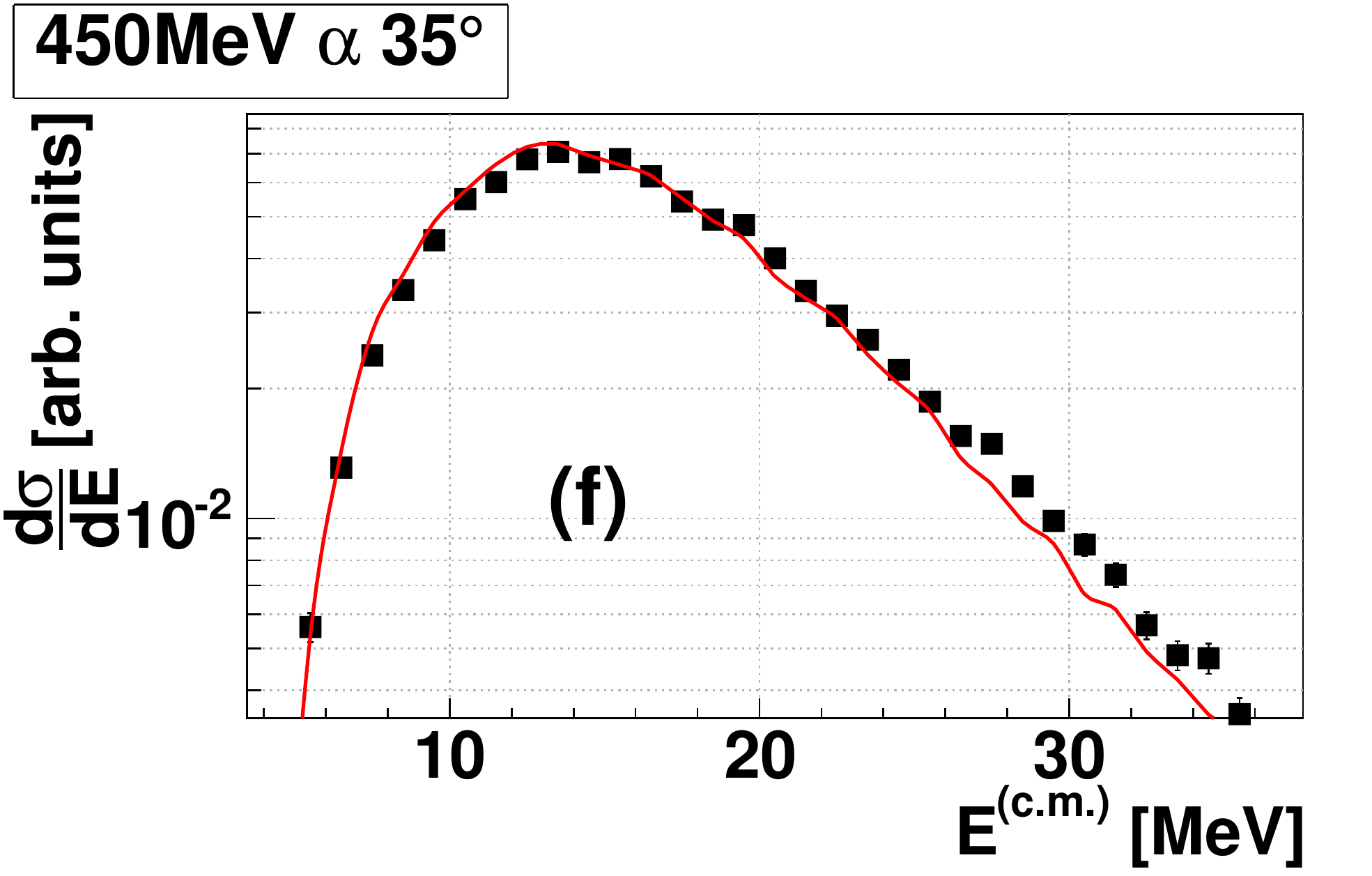}\\
\end{tabular}
\caption{(Color online) The same as Fig.~\ref{fig:ring300} but for the reaction at \SI{450}{MeV}.}
\label{fig:ring450}
\end{figure*}

\begin{figure*}[htpb]
\centering
\begin{tabular}{ccc}
\includegraphics[width=0.3\textwidth] {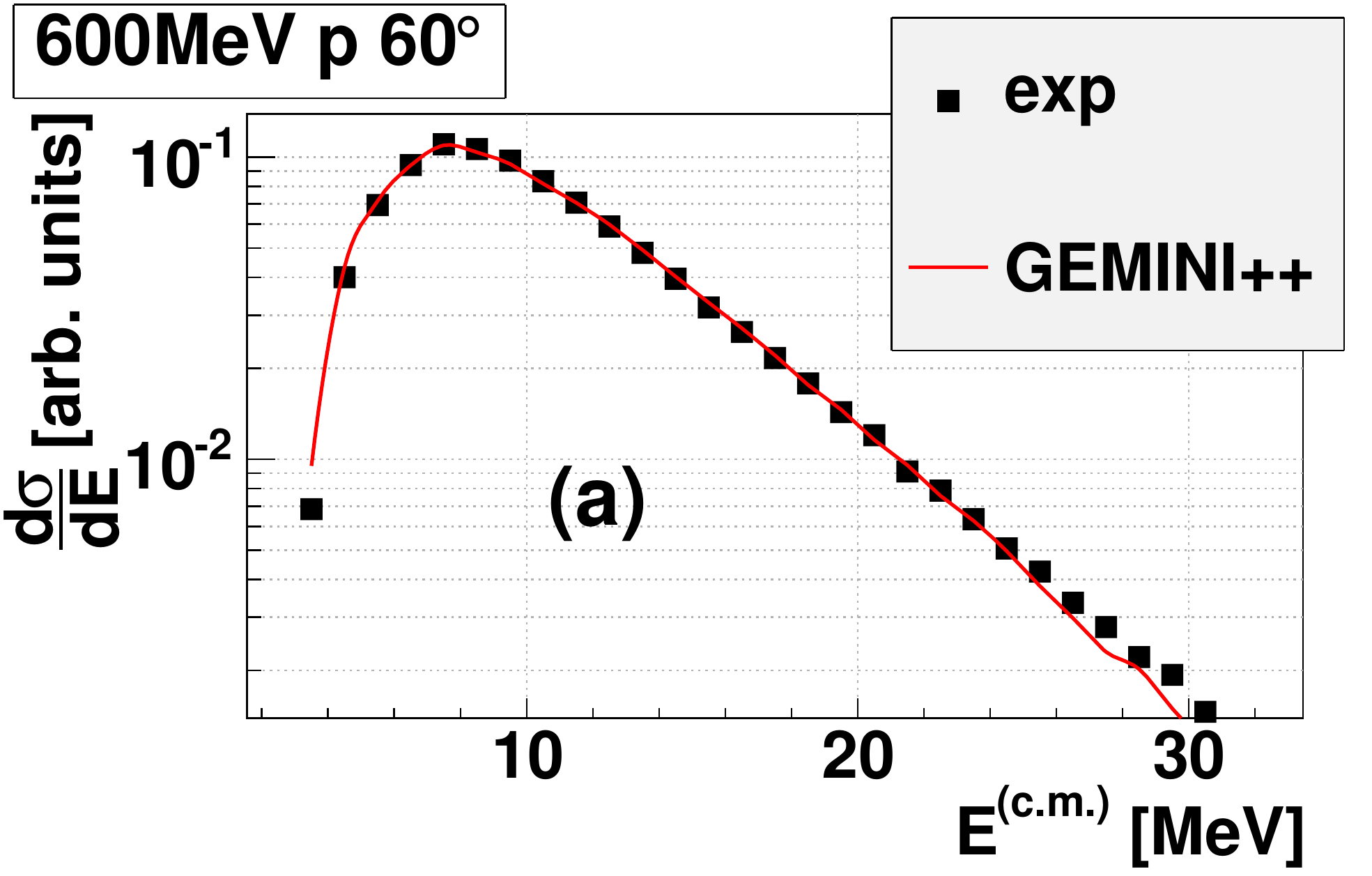} & \includegraphics[width=0.3\textwidth] {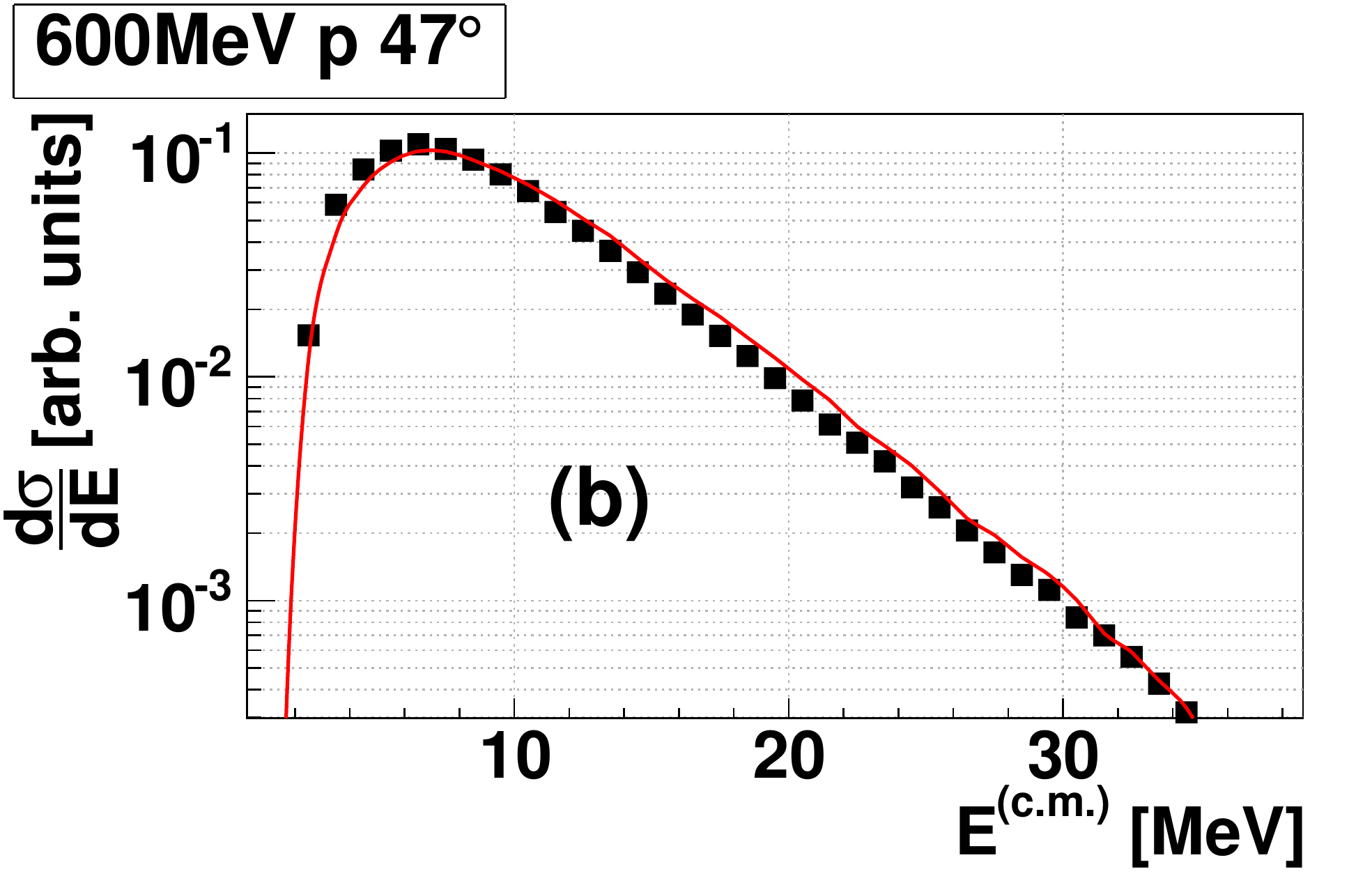} &%
\includegraphics[width=0.3\textwidth] {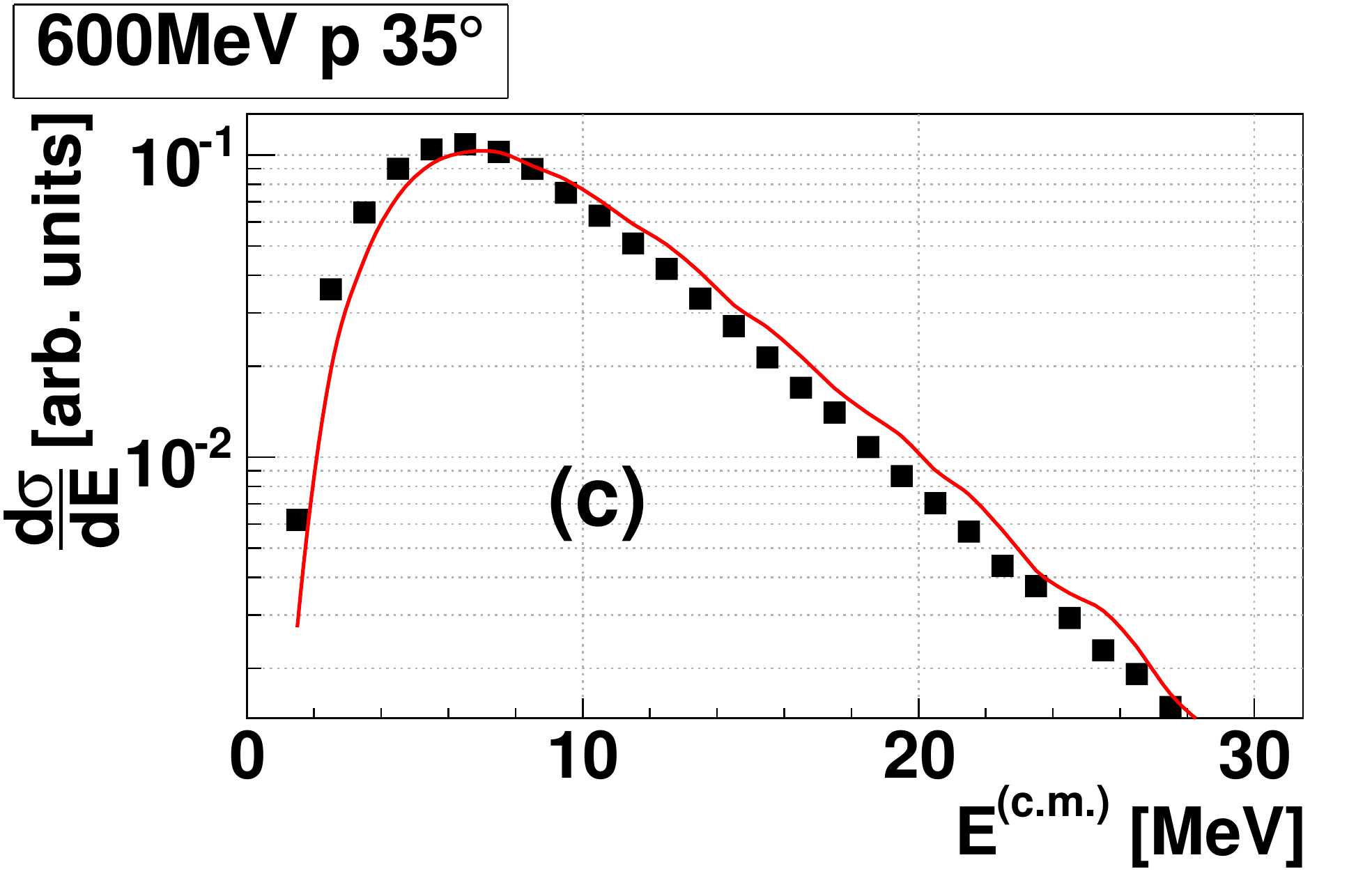}\\
\includegraphics[width=0.3\textwidth] {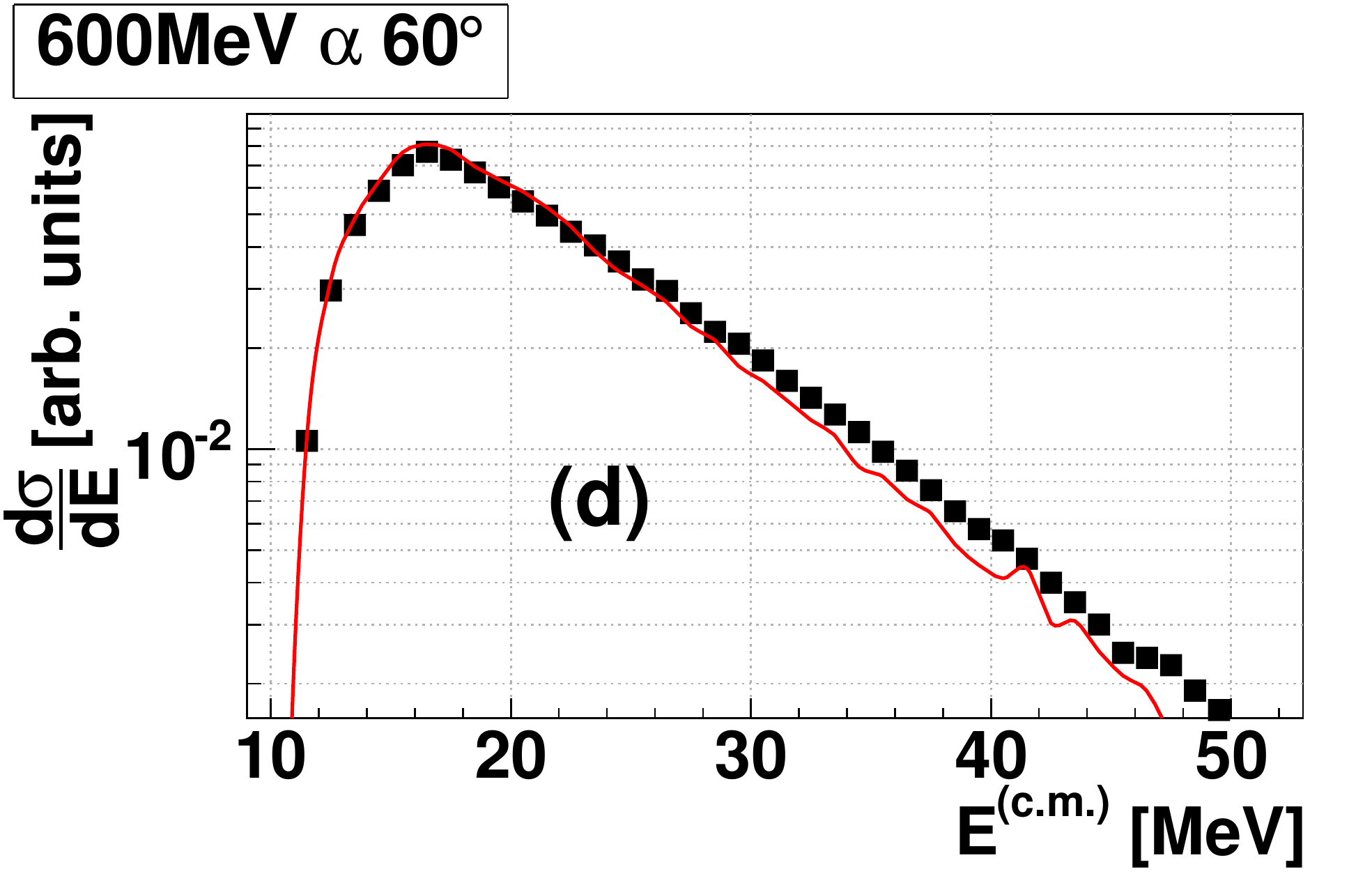} & \includegraphics[width=0.3\textwidth] {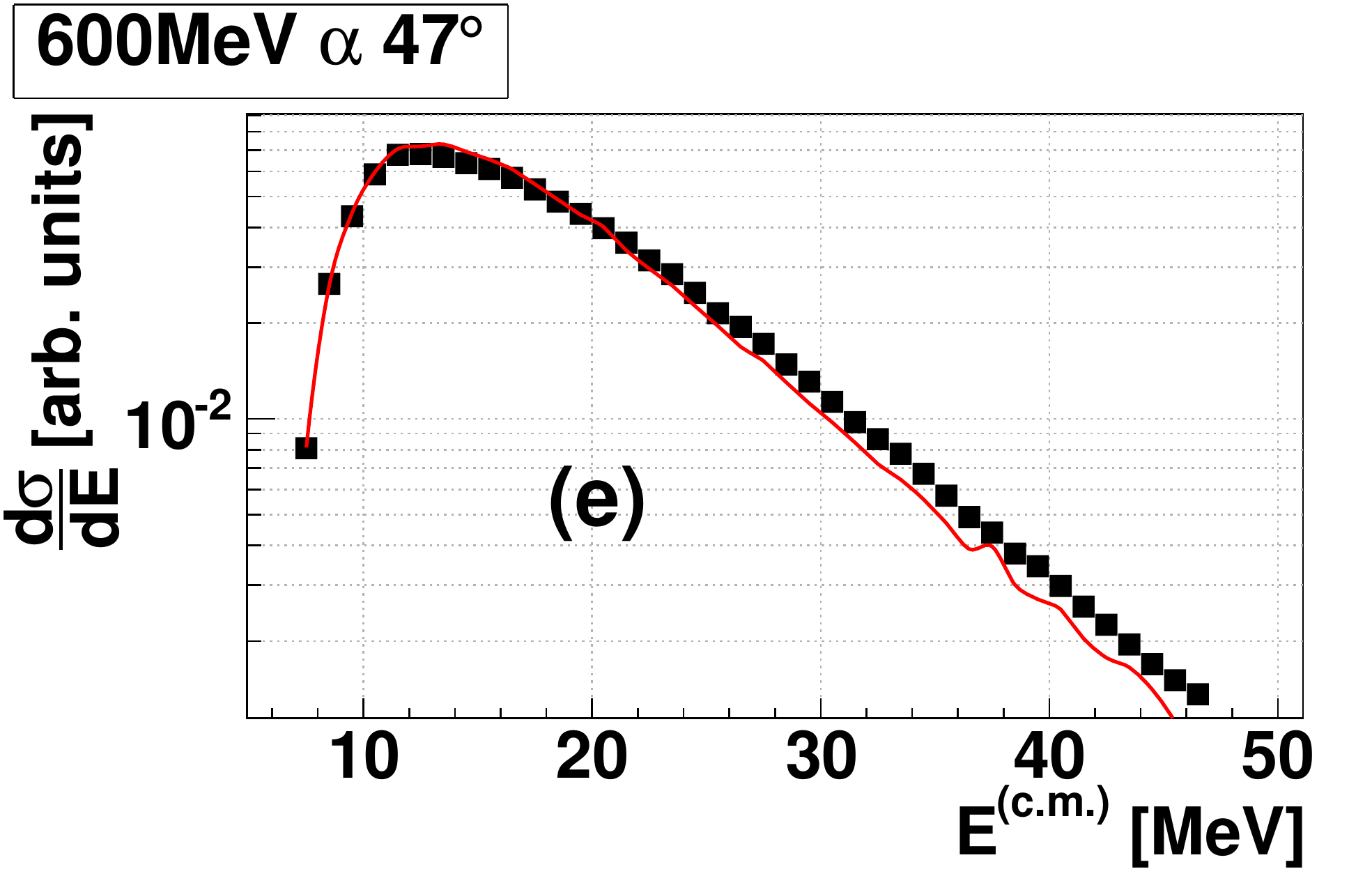} &%
\includegraphics[width=0.3\textwidth] {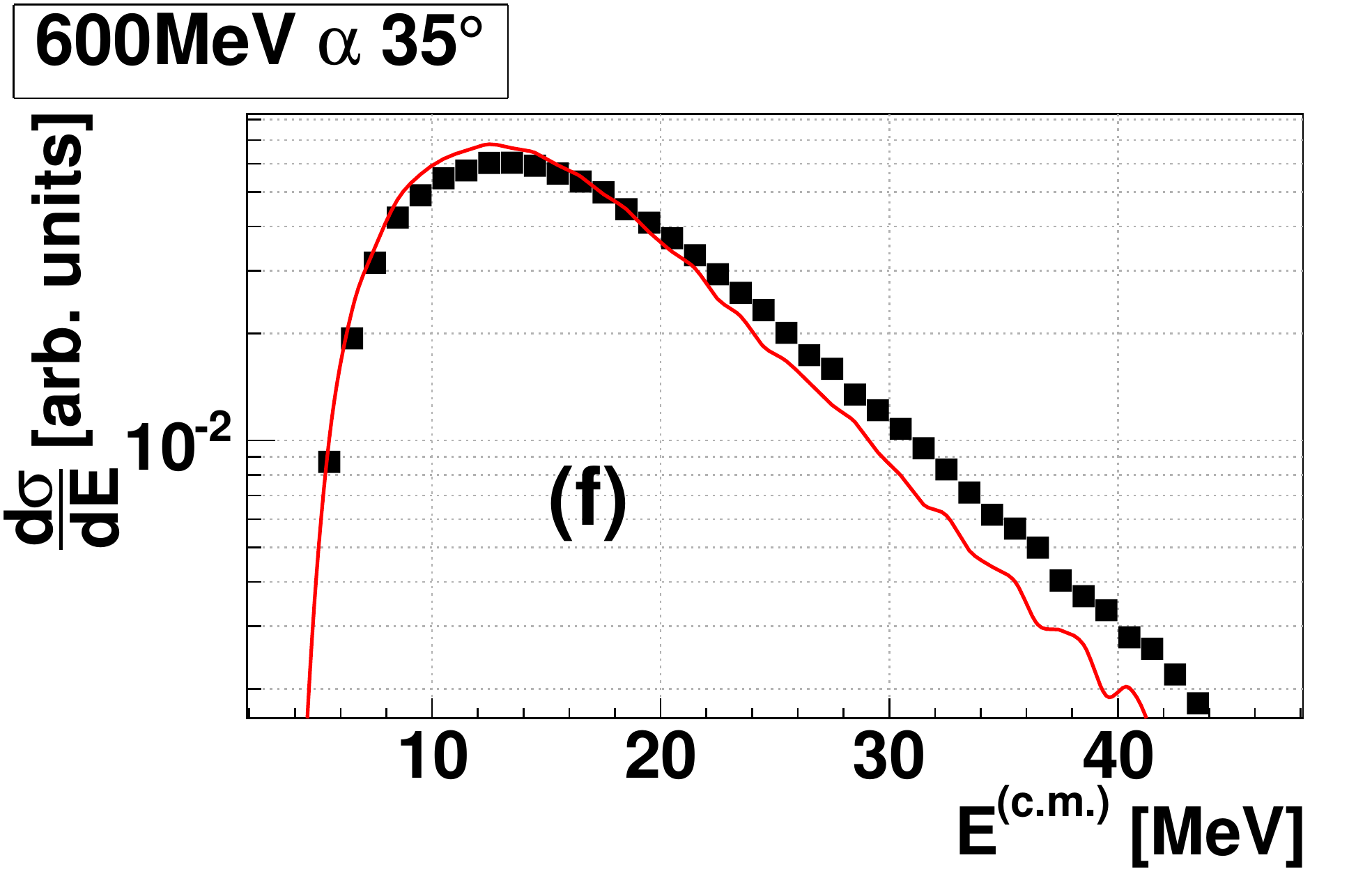}\\
\end{tabular}
\caption{(Color online) The same as Fig.~\ref{fig:ring300} but for the reaction at \SI{600}{MeV}.}
\label{fig:ring600}
\end{figure*}

The center of mass energy spectra of protons (panels (a), (b), (c)) and $\alpha$-particles (panels (d), (e), (f)) for
the reactions at \SI{450}{MeV} and \SI{600}{MeV} are displayed in Fig.~\ref{fig:ring450} and Fig.~\ref{fig:ring600}, respectively.
Also in this case a reasonable agreement at all the explored lab angles is observed between simulated and experimental data.

Laboratory angular distributions for protons (panels (a), (b), (c)) and $\alpha$-particles (panels (d), (e), (f)),
normalized to the number of ER, are shown in Fig.~\ref{fig:angoli} for all the investigated reactions (panels (a) and (d):
\SI{300}{MeV}; (b) and (e): \SI{450}{MeV}; (c) and (f): \SI{600}{MeV}). Experimental data are plotted as black full circles, while
different \gemini\ calculations, all assuming RLDM yrast line and RLDM fission barrier, correspond to different colors.
The horizontal bar in the \garf\ region ($\vartheta >\ang{30}$) indicates the angular coverage of each ring.
The vertical error bar for experimental data represent estimated systematic errors due to the evaluation of the efficiency
in the individual detection cells; statistical errors are negligible.
The main observations are as follows. Concerning proton emission, the data are overall well described by the statistical model
at the three energies, whatever the input parameter choice. There are minor discrepancies which are different for
the three bombarding energies; at \SI{300}{MeV} the model underestimates the proton yields beyond \ang{30}, while
the opposite happens at \SI{600}{MeV}. However the measured proton emission appears to be essentially compatible with
the evaporation from \ce{^{88}Mo} compound nuclei. The most striking difference between experiment and model is for $\alpha$ emission,
as the measured distribution becomes more forward peaked than the model predictions when the energy increases,
almost independently of the fine tuning of the parameters. There are of course some changes if some parameters are modified. For
example, the best set from the point of view of the shape of the energy spectra
($w=\SI{0}{fm}$, $\tau_{d}=10zs$, red squares) gives the worst agreement for the forward emission.
The choice $w=\SI{1.0}{fm}$ (green up triangles and blue down triangles), corresponding to a slightly poorer reproduction
of the $\alpha$ energy spectra (see Fig.~\ref{fig:yrastbfw}), increases the forward focusing of $\alpha$-particles,
with a minor effect of the delay time for fission
($\tau_\mathrm{d}=\SI{0}{zs}$ for blue down triangles, $\tau_\mathrm{d}=\SI{10}{zs}$ for green up triangles).
In any case, the experimental focusing of $\alpha$-particles, increasing with the beam energy,
cannot be properly reproduced by the model code.

\begin{figure}[htbp]
\centering
\includegraphics[width=\columnwidth] {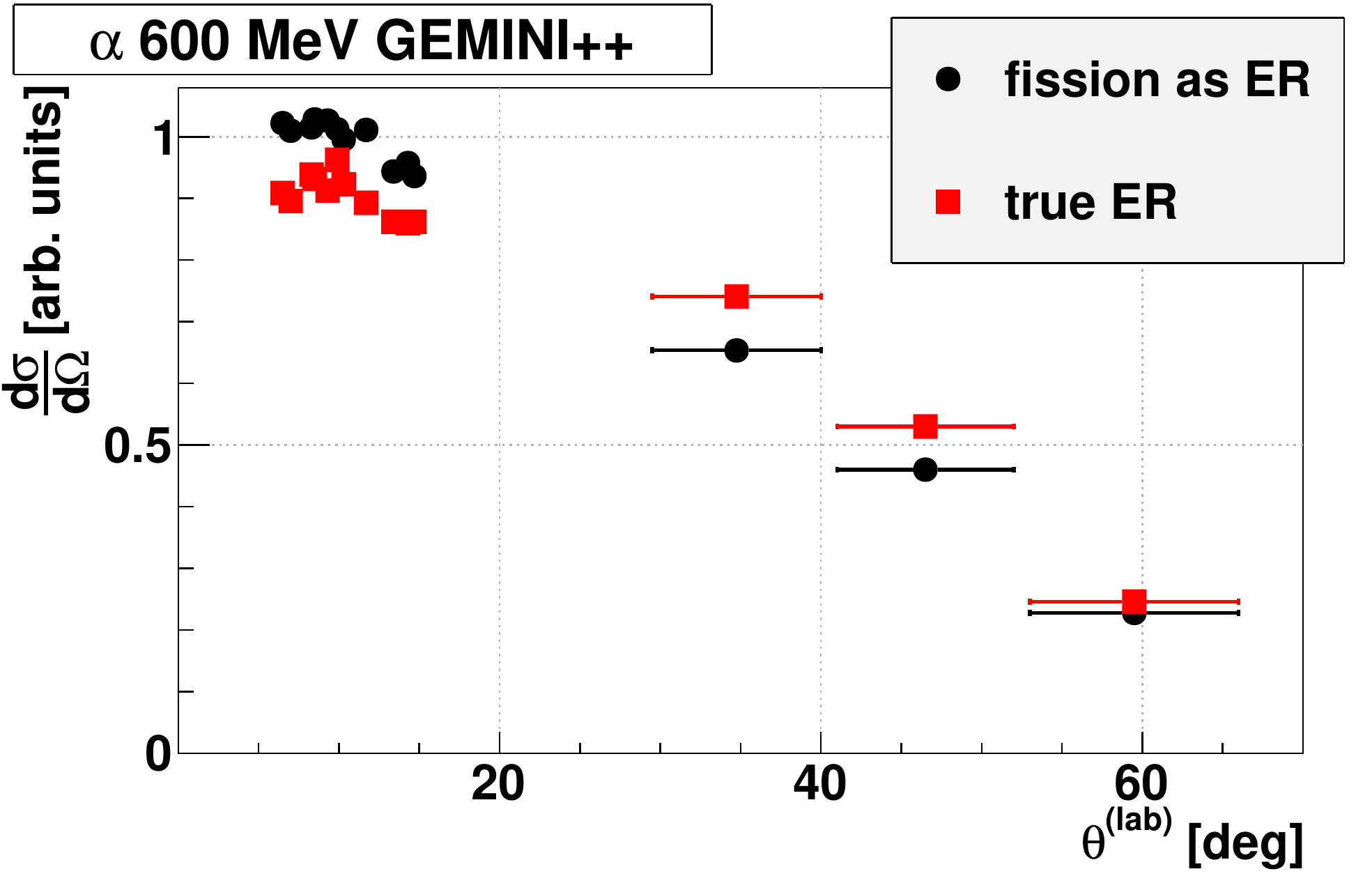}
\caption{
	(Color online) Simulated angular distribution  for $\alpha$-particles  at \SI{600}{MeV};
	\gemini\ data with RLDM yrast line, RLDM fission barrier, $w=\SI{1.0}{fm}$ and $\tau_\mathrm{d}=\SI{0}{zs}$.
	Red squares correspond to events in which a true ER is selected, while black dots correspond to incomplete events
	in which a fission fragment produces a background of the fusion channel.
}
\label{fig:fissionER}
\end{figure}

Concerning the origin of the observed yield discrepancy for $\alpha$-particles we have tested several hypotheses.
For example, the possibility that the \ce{^{40}Ca} target is oxidized (and, as a consequence, causes a spurious emission of $\alpha$ particles)
has been excluded by means of a chemical analysis on the sample. A possible source of $\alpha$ contamination comes from the C layers
embedding the \ce{^{40}Ca}; the detected events can contain a contribution from the complete fusion of \ce{^{48}Ti} and \ce{^{12}C}.
This hypothesis has been rejected too, because of the very different center of mass velocity of the \ce{^{48}Ti + ^{12}C} system
with respect to \ce{^{88}Mo} (they differ by about \SI{12}{mm/ns} at \SI{600}{MeV} and about \SI{9}{mm/ns}
at \SI{300}{MeV}): events coming from the complete fusion of \ce{^{48}Ti} and \ce{^{12}C} are well outside the selected ER gate
(Fig.~\ref{fig:gates}). The fission background of ER events
(not negligible, mainly at \SI{600}{MeV}) is already included in the \gemini\ simulation; any attempt to reduce this spurious
contribution by means of a strict cut on the most forward emitted ER has little effect since \gemini\ shows that the angular
distribution of the background fission fragment is rather broad and flat.
In any case, as shown by Fig.~\ref{fig:fissionER} for $\alpha$-particles at \SI{600}{MeV}, we have verified that,
according to \gemini , particles emitted by a fission fragment erroneously identified as ER are only weakly
more forward focused (full black dots) with respect to the case of true ER (red squares); as a consequence,
they cannot explain the observed discrepancy between simulation and experimental data.

Concerning Deep Inelastic Collisions, not included in the adopted model, we note that for kinematical reasons
only a small tail of QP/QT from very dissipative collisions could fall inside in the ER identification gate.
As a consequence the possible contamination should be negligible, although a simulation of the DIC process
based on a phenomenological parametrization of this mechanism indicates a preferential emission of $\alpha$-particles in the forward direction at \SI{600}{MeV}.
Other processes such as QuasiFusion/QuasiFission are not included in the model and could contaminate in a minor way
the experimental sample of ER events.

\begin{figure}[htbp]
\centering
\includegraphics[width=\columnwidth] {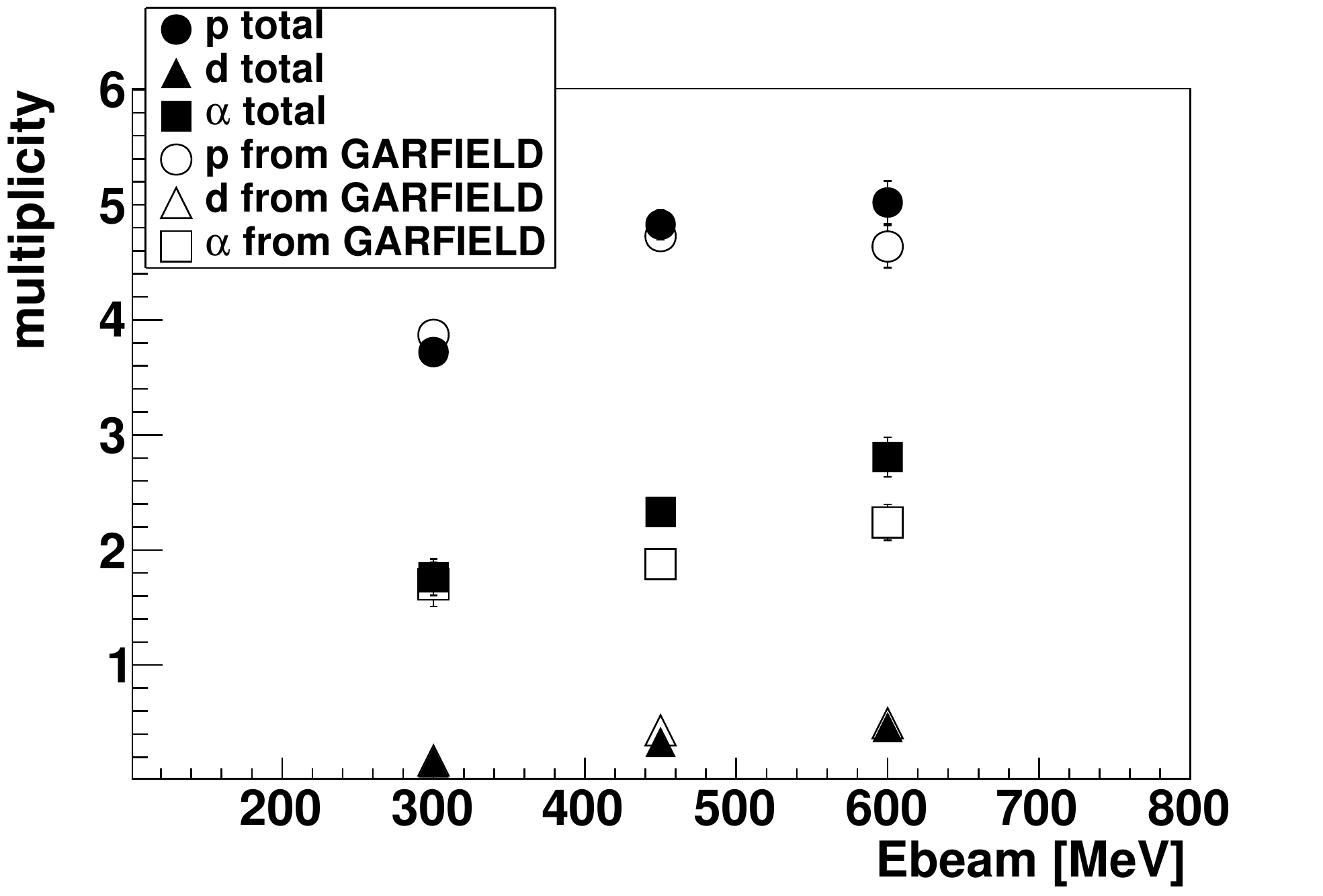}
\caption{
	Estimated experimental particle multiplicities in $4\pi$ as a function of the beam energy in FE events. Circles: protons;
	triangles: deuterons; squares: $\alpha$-particles. Full symbols refer to data from \garf\ and phoswich detectors.
	Open symbols give the particle multiplicities extracted from \garf\ data only.
}
\label{fig:moltepl}
\end{figure}

With the above limitations, we can use \gemini\ data in order to estimate the efficiency for
LCP detection and to correct the measured yields. Various slightly different correction factors have
been deduced according to the different adopted parametrizations of \gemini .
The resulting LCPs $4\pi$-multiplicities obtained by averaging over the values obtained with different parameter
sets are reported in Fig.~\ref{fig:moltepl} for protons, $\alpha$-particles and deuterons, exploiting the isotopic resolution for
hydrogen isotopes both for \garf\ and the phoswich telescopes; tritons are too weak to be reliably evaluated.

Vertical error bars correspond to the standard deviation of the different results.
Each point is also entitled to a statistical error of the order of \SI{8}{\%} for protons (circles), \SI{20}{\%} for deuterons (triangles)
and \SI{10}{\%} for $\alpha$-particles (squares).
The plot shows that particle multiplicities slightly increase with the beam energy, with a weaker trend for protons.
The observed forward focusing of $\alpha$-particles increasing with bombarding energy suggests the
onset of pre-equilibrium processes. Therefore, we tried to estimate $\alpha$ multiplicities separately
from \garf\ data only (extrapolating them to the full solid angle),
where angular distributions and yields are quite well reproduced by the statistical model.
The same choice can be applied also to protons, although no clear evidence of a model failure with increasing energy
exists (Fig.~\ref{fig:angoli} top part). The obtained results are shown as open symbols in the same Fig.~\ref{fig:moltepl}. For
$\alpha$-particles the difference between the two data sets represents an upper limit for pre-equilibrium emission that,
as expected, increases with the beam energy, or, in any case, for emission sources not included in the simulation.
On the contrary, for protons there is no significant discrepancy at all energies; even at \SI{600}{MeV} the two extracted
proton multiplicities are compatible within the errors. For $\alpha$-particles the possible pre-equilibrium
contribution starts at \SI{450}{MeV} and corresponds to about 0.5 particles per event
(with respect to a total number of 2.8 $\alpha$-particles per event) in the worst case (\SI{600}{MeV}),
thus justifying the assumption of negligible pre-equilibrium effects done in \cite{Ciemala15} for the evaluation of the GDR strength.

By means of the fast plastic scintillator located below the grazing angle for all reactions
(see section~\ref{sec:experiment}) it has been possible to estimate absolute cross sections.
For FE events, the obtained values are reported in Table~\ref{tab:cross} in the row labelled $\sigma_\mathrm{FE}$;
they are the average among the various estimated $\sigma_\mathrm{FE}$ values obtained running \gemini\ with
different sets of parameters (all of them with RLDM yrast line and RLDM fission barrier). The error
is the standard deviation of these values. It is interesting to note that, as shown in the last row of the table,
the FE cross section is almost one half of the reaction cross section (last column of Table~\ref{tab:param})
at \SI{300}{MeV}, while it decreases to about \SI{19}{\%} at \SI{600}{MeV}.

\begin{table}[htpb]
\caption{
	Experimental absolute cross sections. $\sigma_\mathrm{FE}$ is the fusion-evaporation cross section; $\sigma_\mathrm{FF}$ is
	the fusion-fission cross section. $\sigma_\mathrm{F} / \sigma_\mathrm{R}$ is the ratio between the
	total fusion cross section (FF plus FE) and the total reaction cross section,
	taken from the last column of Table~\ref{tab:param}. $\sigma_\mathrm{FE} / \sigma_\mathrm{R}$ is the ratio between the
	FE cross section and the total reaction cross section.
}
\centering
\begin{tabular}{cccc}
\toprule
\T\B & \bf 300$\,$MeV & \bf 450$\,$MeV & \bf 600$\,$MeV\\
\colrule
\T $\sigma_\mathrm{FE}$                     & $(893\pm 109)\,$mb & $(545\pm45)\,$mb & $(459\pm115)\,$mb\\
$\sigma_\mathrm{FF}$                        & $(115\pm3)\,$mb    & $(266\pm37)\,$mb & $(417\pm114)\,$mb\\
$\sigma_\mathrm{F} / \sigma_\mathrm{R}$     & $0.54\pm0.06$      & $0.358\pm0.004$  & $0.36\pm0.07$\\
\B $\sigma_\mathrm{FE} / \sigma_\mathrm{R}$ & $0.48\pm0.06$      & $0.24\pm0.02$    & $0.19\pm0.05$\\
\botrule
\end{tabular}
\label{tab:cross}
\end{table}

\begin{figure}[htbp]
\centering
\includegraphics[width=\columnwidth] {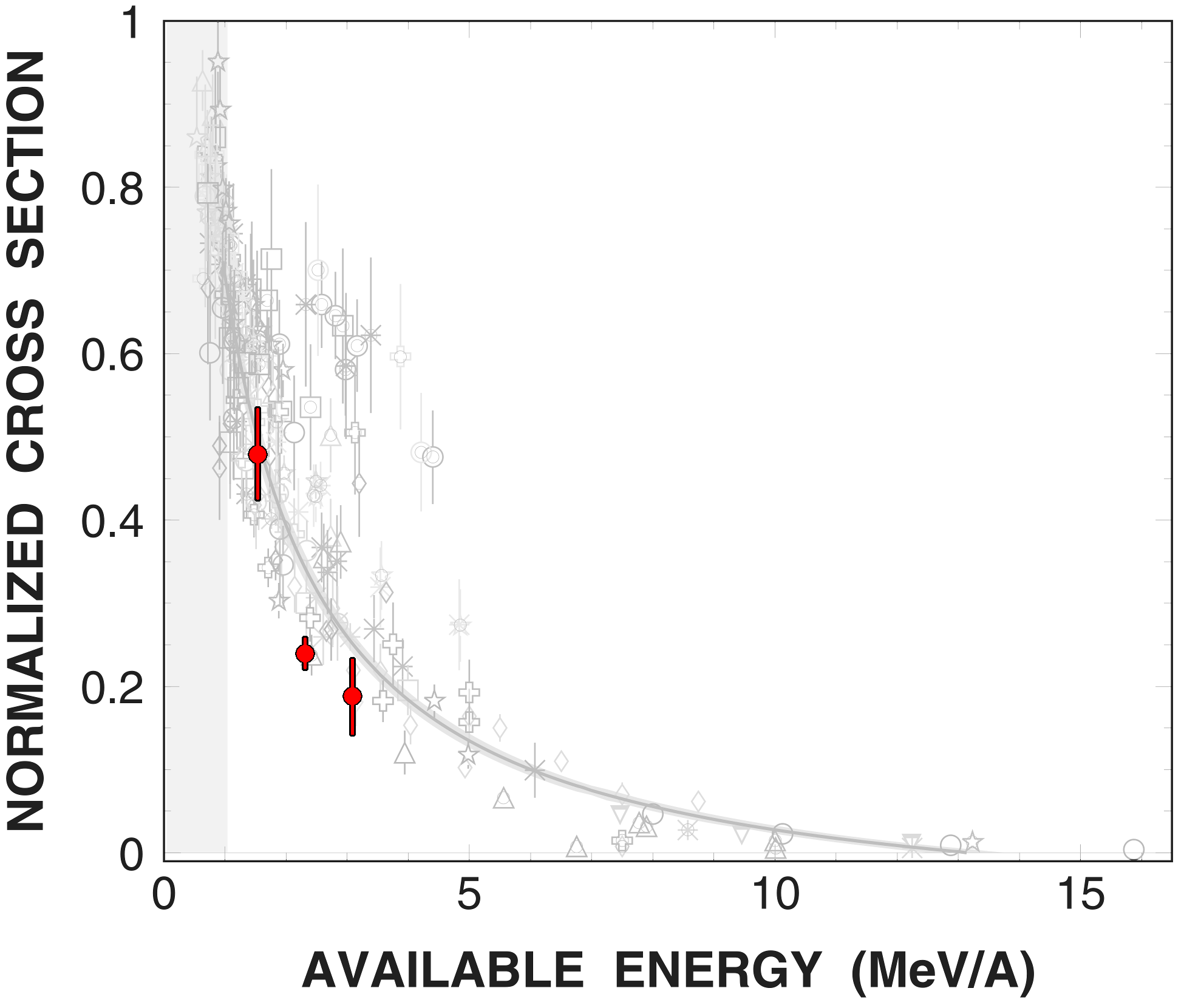}
\caption{
	(Color online) FE cross section normalized to the reaction cross section as a function of the $E_\mathrm{cm}/\mathrm{nucleon}$.
	Red dots correspond to the experimental values found in this work, compared with the systematics of \cite{Eudes13}.
	The figure is adapted from Fig.~2 in \cite{Eudes13}.
}
\label{fig:eudes}
\end{figure}

If we plot the normalized FE cross section $\sigma_\mathrm{FE} / \sigma_\mathrm{R}$ as a function of the available c.m.
energy per nucleon as proposed in the recent systematics for fusion evaporation \cite{Eudes13},
we can see (Fig.~\ref{fig:eudes}) that our data (red dots) fairly agree with the reported prescription.

Finally, we briefly discuss the fission case in order to extract the absolute cross section for this channel.
The fission process is mainly associated with the highest angular momenta, as it is evident from Fig.~\ref{fig:byrast},
dotted lines. In the adopted parametrization (RLDM fission barrier) the barrier is around \SI{50}{MeV} at low $J$
(as calculated also in \cite{Dobrowolski09}) and decreases smoothly, until vanishes around \SI{79}{\planckbar}.
In the experimental data the symmetric fission is weak, the measured fraction varying from \SI{10}{\%} to
\SI{30}{\%} of the total fission events (without geometrical correction), depending on the bombarding energy.
The absolute fission cross section $\sigma_\mathrm{FF}$ is reported in Table~\ref{tab:cross} second row:
its weight in the total fusion cross section smoothly increases from \SI{300}{MeV} to \SI{600}{MeV}.
At \SI{600}{MeV} the FF cross section is of the same order of magnitude as the FE cross section.
If we compare the total fusion cross section with the total reaction cross section $\sigma_\mathrm{F} / \sigma_\mathrm{R}$
(third line of Table~\ref{tab:cross}) we can see that also at \SI{300}{MeV} there is room for other reaction mechanisms
(such as DIC): in fact fusion represents only \SI{54}{\%} of the total reaction cross section.

\section{Summary and conclusions}
In this work we presented some exclusive data concerning the system \ce{^{48}Ti + ^{40}Ca} at three bombarding energies
(\SIlist{300;450;600}{MeV}). Data have been collected by means of the \garf\ setup
coupled to a wall of phoswich telescopes of the \textsc{Fiasco} device.

We have focused our analysis on the fusion channel whose collective dipolar excitation
has been the subject of our recent investigation \cite{Ciemala15}. In particular we have shown that:
\begin{itemize}
\item this channel represents more than \SI{50}{\%} of the total reaction cross section at \SI{300}{MeV}
and it goes down to about \SI{40}{\%} at \SI{600}{MeV}.
\end{itemize}

As expected, the formed CN decays through the evaporation of LCPs (fusion-evaporation event) or can undergo fission (fu\-sion-fission event).
After selecting FE events, we have constrained the main parameters of the statistical code \gemini\ comparing
measured and simulated LCP distributions at \SI{300}{MeV} where contributions from sources other than the CN should be negligible.
We have seen that:
\begin{itemize}
\item proton energy spectra are reproduced by \gemini\ with the standard parameters;
\item in the case of $\alpha$-particles it is necessary to use the RLDM parametrization
both for the yrast line and the fission barrier, and to reduce the LCP Coulomb barrier fluctuations.
This can indicate rather spherical excited nuclei as the sources of the detected LCP;
\item at the two highest beam energies there is an excess of $\alpha$-particles at forward angles,
which cannot be reproduced by the statistical model;
\item this excess, corresponding on the whole to about \SI{20}{\%} of the total emitted $\alpha$-particles
at the highest energy (\SI{600}{MeV}), may be due both to pre-equilibrium emission and to other processes
(such as Deep Inelastic Collision, QuasiFission/QuasiFusion), whose residual contamination cannot be definitely excluded
due to the difficulty of a very clean fusion channel selection;
\item this excess of $\alpha$ emission can be taken as an upper limit for the pre-equilibrium emission of this system,
thus justifying the assumption done in \cite{Ciemala15} of negligible pre-equilibrium.
\end{itemize}

Finally we have extracted the absolute cross section for the FE and  FF cases:
\begin{itemize}
\item the FE cross sections are in reasonable agreement with the existing systematics reported in \cite{Eudes13};
\item the total fusion cross section is well below the total reaction cross section
(estimated according to \cite{Gupta84}), suggesting the presence of other processes, such as DIC, sizably
contributing at the highest angular momenta.
\end{itemize}

The authors thank professor P. R. Maurenzig, retired from Dipartimento di Fisica,
Università degli Studi di Firenze, Italy, for the useful discussions during the data analysis presented in this work.

The authors would like to thank also Prof. R. J. Charity from Washington University of St. Louis (Missouri, USA)
for the support in the use of the \gemini\ code.

This work was possible thanks to the good beam provided by LNL accelerator crew and
to the work of the LNL target lab staff.

This work was supported by the Polish Ministry of Science and Higher Education, Grants No. N N202 486339,
No. 2011/03/B/ST2/01894, and No. 2013/08/M/ST2/00591 and by
the French-Polish agreements IN2P3-COPIN (Projects No. 06-126, No. 05-119, No. 09-136, and No. 12-145).

This work was also partially supported by grants of Italian Ministry of Education, University and
Research under contract PRIN 2010-2011.

\bibliography{biblio}

\end{document}